\documentclass[sigconf,screen]{acmart}

\acmPrice{15.00}
\acmDOI{10.1145/3519935.3519977}
\acmYear{2022}
\copyrightyear{2022}
\acmSubmissionID{stoc22main-p153-p}
\acmISBN{978-1-4503-9264-8/22/06}
\acmConference[STOC '22]{Proceedings of the 54rd Annual ACM SIGACT Symposium on Theory of Computing}{June 20--24, 2022}{Rome, Italy}
\acmBooktitle{Proceedings of the 54rd Annual ACM SIGACT Symposium on Theory of Computing (STOC '22), June 20--24, 2022, Rome, Italy}

\usepackage{jeffe} 
\DeclareMathAlphabet{\mathcal}{OMS}{cmsy}{m}{n}

\usepackage{graphicx}
\graphicspath{{figures/}}
\DeclareGraphicsExtensions{.pdf,.png,.jpg}

\usepackage{mathtools}
\usepackage{enumerate}
\usepackage{enumitem}
\setitemize{itemsep=0pt}
\setenumerate{itemsep=0pt}

\def\FullVersion{1}

\ifx\FullVersion\undefined
   \newcommand{\FullVer}[1]{}
   \newcommand{\NotFullVer}[1]{#1}
\else
   \newcommand{\FullVer}[1]{#1}
   \newcommand{\NotFullVer}[1]{}
\fi

\makeatletter
\long\def\@makecaption#1#2{
    \vskip 10pt
    \setbox\@tempboxa\hbox{{\footnotesize {\bf #1.} #2}}
    \ifdim \wd\@tempboxa >\hsize         
        {\footnotesize {\bf #1.} #2\par}
      \else                              
        \hbox to\hsize{\hfil\box\@tempboxa\hfil}
    \fi}
\makeatother

\def\mypara#1{\smallskip\noindent\textbf{#1} \,\,}

\usepackage[color=red, backgroundcolor=TODOcolor, textcolor=red, linecolor=lightgray, bordercolor=red, textsize=scriptsize]{todonotes}

\renewcommand{\note}[2][] {%
\todo[#1]
    { %
        \textsf{\boldmath\textbf{#2}} \leavevmode
    } %
}

\def\etal{\textit{et~al.}}
\def\polylog{\mathop{\mathrm{polylog}}}
\def\eps{\varepsilon}

\def\reals{\mathbb{R}}
\def\ints{\mathbb{Z}}

\usepackage{textcomp}
\DeclareRobustCommand{\cost}{\textit{\raisebox{0.27pt}{\scalebox{0.97}{\textcent}}}}

\def\diam{\operatorname{diam}}

\def\net#1{\bar{#1}}

\def\adj{\alpha}
\def\pnlty{\delta}
\def\cnter{\operatorname{cntr}} 
\def\Pairs{\textit{OptPairs}}
\def\canonical{\mathsf{C}}

\usepackage{accents}

\DeclareRobustCommand\ubar[1]{%
\underaccent{\bar}{#1}}

\makeatletter
\def\fdsy@scale{1}
\newcommand\fdsy@mweight@normal{Book}
\newcommand\fdsy@mweight@small{Book}
\newcommand\fdsy@bweight@normal{Medium}
\newcommand\fdsy@bweight@small{Medium}
\DeclareFontFamily{U}{FdSymbolA}{}
\DeclareFontShape{U}{FdSymbolA}{m}{n}{
    <-7.1> s * [\fdsy@scale] FdSymbolA-\fdsy@mweight@small
    <7.1-> s * [\fdsy@scale] FdSymbolA-\fdsy@mweight@normal
}{}
\DeclareFontShape{U}{FdSymbolA}{b}{n}{
    <-7.1> s * [\fdsy@scale] FdSymbolA-\fdsy@bweight@small
    <7.1-> s * [\fdsy@scale] FdSymbolA-\fdsy@bweight@normal
}{}
\makeatother

\DeclareSymbolFont{fdops}{U}{FdSymbolA}{m}{n}
\SetSymbolFont{fdops}{bold}{U}{FdSymbolA}{m}{n}

\DeclareMathSymbol{\smalltriangleright}{\mathrel}{fdops}{78}
\DeclareMathSymbol{\smalltriangleleft}{\mathrel}{fdops}{80}
\DeclareMathSymbol{\smallblacktriangleright}{\mathrel}{fdops}{82}
\DeclareMathSymbol{\smallblacktriangleleft}{\mathrel}{fdops}{84}
\def\spL{\smallblacktriangleright}
\def\spR{\smallblacktriangleleft}
\def\spoL{\smalltriangleright}
\def\spoR{\smalltriangleleft}

\def\cell{\boxplus}
\def\subcell{\square}

\def\Grid{\mathfrak{G}}

\def\Cover{\mathcal{C}}
\def\childr{\operatorname{Ch}}

\def\level{\operatorname{level}}

\def\opt{{\mathrm{opt}}}
\def\alg{{\mathrm{alg}}}

\def\C{\mathcal{C}}

\definecolor{BrickRed}{rgb}{.72,0,0} 
\def\EMPH#1{\emph{\textcolor{BrickRed} {#1}}}

\newcommand{\GLOS}[3][]{%
\EMPH{\ensuremath{#3}}%
}

\begin{document}

\title{Deterministic, Near-Linear $\varepsilon$-Approximation Algorithm for Geometric Bipartite Matching%
}

\author{Pankaj K.\ Agarwal}
\affiliation{
    \institution{Duke University}
    \country{USA}
}

\author{Hsien-Chih Chang}
\affiliation{
    \institution{Dartmouth College}
    \country{USA}
}

\author{Sharath Raghvendra}
\affiliation{
    \institution{Virginia Tech}
    \country{USA}
}

\author{Allen Xiao}
\affiliation{
    \institution{Duke University}
    \country{USA}
}


\date{\today}

\begin{abstract}
Given point sets $A$ and $B$ in $\mathbb{R}^d$ where $A$ and $B$ have equal size $n$ for some constant dimension $d$ and a parameter $\varepsilon>0$, we present the first deterministic algorithm that computes, in $n\cdot(\varepsilon^{-1} \log n)^{O(d)}$ time, a perfect matching between $A$ and $B$ whose cost is within a $(1+\varepsilon)$ factor of the optimal under any $\smash{\ell_p}$-norm.
Although a Monte-Carlo algorithm with a similar running time is proposed by Raghvendra and Agarwal~[J.\ ACM 2020], the best-known deterministic $\varepsilon$-approximation algorithm takes $\Omega(n^{3/2})$ time. Our algorithm constructs a (refinement of a) tree cover of $\mathbb{R}^d$, and we develop several new tools to apply a tree-cover based approach to compute an $\varepsilon$-approximate perfect matching.
\end{abstract}

\maketitle



\section{Introduction}

Let $A$ and $B$ be two point sets in $\reals^d$ of size $n$ each, where the dimension $d$ is a constant.
Consider the complete weighted bipartite graph $G$ with cost function
$\EMPH{$\cost(e)$} \coloneqq \norm{a-b}$, where $\norm{\cdot}$ denotes the Euclidean norm.\footnote{Our algorithm works for any $\ell_p$-norm, but for the sake of concreteness of the presentation we use the $\ell_2$-norm.}
A \EMPH{matching} $M$ is a set of vertex-disjoint edges in $G$.
We say that a matching is \EMPH{perfect} if $\abs{M}=n$.
The cost of $M$ is the sum of its edge costs:
\(
\cost(M) \coloneqq \sum_{e \in M} \cost(e).
\)
The \EMPH{Euclidean minimum-weight matching} (EMWM) in $G$ is denoted as
\(
\GLOS[opt-matching]{Minimum-weight perfect matching}{M_\opt} \coloneqq \argmin_{\abs{M} = n} \cost(M).
\)
A perfect matching $M$ is called an \EMPH{$\eps$-approximation} if $\cost(M) \le (1+\eps) \cdot \cost(M_\opt)$.

The EMWM can be used to estimate the Wasserstein distance (a measure of similarity) between two continuous probability distributions. 
\FullVer{
Due to this, it has received considerable attention in machine learning and computer vision~\cite{rt_ijcv00,solomon2015convolutional,PC19,wgan}.
}%
\NotFullVer{
Due to this, it has received considerable attention in machine learning and computer vision~\cite{rt_ijcv00,solomon2015convolutional,PC19}.
}
\FullVer{
For instance, suppose we are given two (possibly unknown) continuous distributions in $\reals^d$, we can estimate their Wasserstein distance by taking $n$
samples from each distribution and then computing a minimum-weight matching between them~\cite{wgan,WB17,Sol18,beugnot2021improving,pmlr-v80-liu18d}.
}%
\NotFullVer{
For instance, suppose we are given two (possibly unknown) continuous distributions in $\reals^d$, we can estimate their Wasserstein distance by taking $n$
samples from each distribution and then computing a minimum-weight matching between them~\cite{beugnot2021improving,pmlr-v80-liu18d}.
}
These applications call for the design of exact and approximation algorithms for the EMWM problem.
In this paper, we present a \emph{deterministic} near-linear-time $\eps$-approximation algorithm for the EMWM problem.
All known near-linear-time algorithms for this problem are Monte-Carlo algorithms, and all existing deterministic $\eps$-approximations take $\Omega(n^{3/2})$ time.

\paragraph{Related work.}
The classical Hopcroft-Karp algorithm computes a maximum-cardinality matching in a bipartite graph with $n$ vertices and $m$ edges in $O(m\sqrt{n})$ time~\cite{hk_sicomp73}.
The first improvement in over thirty years, by Mądry~\cite{madry2013navigating}, runs in $O(m^{10/7} \polylog n)$ time.
\FullVer{%
The bound was further improved to $O((m+n^{3/2})\polylog n)$ by Brand~\etal~\cite{Brand20} (see also Kathuria-Liu-Sidford~\cite{kls-ucma-2020}).}
\NotFullVer{%
The bound was further improved to $O((m+n^{3/2})\polylog n)$ by Brand~\etal~\cite{Brand20}.}
The Hungarian algorithm computes the minimum-weight maximum cardinality matching in $O(mn+n^2\log n)$ time~\cite{p_book}; see also~\cite{ft-fhtui-1987,gt-fsanp-1989}. 
\FullVer{%
Faster algorithms for computing optimal matchings can be obtained by using recent min-cost max-flow algorithms~\cite{bll+-mcfm-2021}.
There is also extensive work on computing optimal matchings and maximum-cost matchings in non-bipartite graphs~\cite{dp_focs10,gt_jacm91,mv_focs80,gab-wmamc-2017,vaz-pmma-2020,dps-sawmg-2018}.%
}
\NotFullVer{%
Faster algorithms for computing optimal matchings can be obtained by using recent min-cost max-flow algorithms~\cite{Brand20}.
There is also extensive work on computing optimal matchings matchings in non-bipartite graphs~\cite{gt_jacm91,gab-wmamc-2017,vaz-pmma-2020}.}

If $A$ and $B$ are points in $\reals^2$,
the best known algorithm for computing EMWM runs in $O(n^2 \polylog n)$ time~\cite{aes-vdsl3-2000,kmr+-dpvdg-2017,acx-eagpm-2019}.
If points have integer coordinates bounded by $\Delta$, the running time can be improved to $O(n^{3/2} \polylog n \log \Delta)$~\cite{s_socg13}.
It is an open question whether a subquadratic algorithm exists for computing EMWM if coordinates of input points have real values.
In contrast, Varadarajan~\cite{v_focs98} presented an $O(n^{3/2}\polylog n)$-time algorithm for the non-bipartite case under any $\ell_p$-norm --- this is surprising because the non-bipartite case seems harder for graphs with arbitrary edge costs.
%
%
As for higher dimensions,
Varadarajan and Agarwal~\cite{VaradarajanA99} presented an $\smash{O(n^{3/2}\eps^{-d}\log^d n)}$-time {$\eps$-approximation} algorithm for computing EMWM of points lying in $\reals^d$.
The running time was later improved to $O(n^{3/2}\eps^{-d}\log^5 n)$ by Agarwal and Raghvendra~\cite{sa_soda12}.
For any $0 < \delta \le 1$, they also proposed a deterministic $O(1/\delta)$-approximation algorithm that runs in $O(n^{1+\delta}\log^d n)$ time~\cite{AgarwalS14}.

Randomly-shifted quadtrees have played a central role in designing Monte-Carlo approximation algorithm for EMWM.
It is well-known that a simple greedy algorithm on a randomly-shifted quadtree yields (in expectation) an $O(\log n)$-approximation algorithm of EMWM~\cite{Ch02}.
Agarwal and Varadarajan~\cite{av-ncaeb-2004} build upon this observation and use a randomly-shifted-quadtree-type hierarchical structure to obtain an expected $O(\log 1/\delta)$-approximation of EMWM in $O(n^{1+\delta})$ time.
Combining their approach with importance sampling, Indyk~\cite{ind-ltcfa-2007} presented an algorithm that approximates the cost of EMWM within an $O(1)$-factor with high probability in time $O(n\polylog n)$.
His algorithm, however, only returns the optimal cost but not the matching itself.
Similarly,
Andoni~\etal~\cite{anoy-paggp-2014} gave an $\e$-approximation streaming algorithm to the cost that runs in $O(n^{1+o_\e(1)})$ time.
Finally, Raghvendra and Agarwal~\cite{ra-ntag-2020} proposed a Monte-Carlo algorithm that computes an $\e$-approximation with high probability in $n\cdot(\eps^{-1}\log n)^{O(d)}$ time.
Roughly speaking, their algorithm embeds the Euclidean distance into a refinement of a randomly-shifted quadtree $T$
that, in expectation, $\eps$-approximates the Euclidean distance.  It iteratively computes the minimum net-cost augmenting path with
respect to this distance and augments the matching along the path. 
To compute the minimum net-cost path, it only stores $O(\mathrm{poly}\log n)$ sub-paths (alternating paths) at each cell of the quadtree. They show that any sub-path at a cell, including the minimum net-cost augmenting path, can be expressed as some combination of the $O(\mathrm{poly}\log n)$ sub-paths stored at its children. 
Using this observation, they build a dynamic data structure that stores a matching, and supports computing and augmenting along a minimum net-cost path in time proportional to its length.  Similar to the Gabow-Tarjan algorithm~\cite{gt-fsanp-1989}, a small fixed penalty is added to the net cost of each edge
so that
the total length of the augmenting paths computed by the algorithm is $O(\e^{-1} n \log n)$, leading to a near-linear-time $\eps$-approximation algorithm.
The randomized quadtree framework also has been successfully applied to designing fast approximation algorithm for the 
transportation problem as well as for matching under different cost 
functions~\cite{preconditioningTransport, fox-lu-transport, afpvx17,lr-tar-2021}.  


\bigskip
\noindent\textbf{Our results.} \,\,
The following theorem
is our main result:
\begin{theorem}
\label{Th:main_alg}
Let $A$ and $B$ be two point sets in $\reals^d$ of size $n$ each, where dimension $d$ is a constant, and let $\eps > 0$ be a parameter.
A perfect matching of $A$ and $B$ of cost at most $\smash{(1+\eps)\cdot \cost(M_\opt)}$ can be computed in $\smash{n \cdot (\e^{-1}\log n})^{O(d)}$ time in the worst case.
\end{theorem}


\noindent If we try to design a deterministic algorithm for EMWM using an approach similar to Raghvendra and Agarwal~\cite{ra-ntag-2020}, we run into several difficulties.
Neither can we use a randomly-shifted quadtree, nor can we use a stochastic embedding of Euclidean metric into a tree metric.
Instead we work with Euclidean distance directly.
We replace a single randomly-shifted quadtree with $2^d$ deterministically shifted quadtrees and then define a refinement on each of
these trees. Our shifted quadtrees can be viewed as a \emph{tree cover} of Euclidean space, a notion introduced by Gupta, Kumar, and Rastogi~\cite{gkr-tpdr-2005};
the Euclidean distance between a pair of points is deterministically approximated in at least one of these $2^d$ quadtrees.
\FullVer{%
The previous applications of tree covers were mostly limited to \emph{decomposable problems} where one can solve the problem by simply merging
the solutions for all the trees, for example like routing, spanners, nearest-neighbor searching, and network design~\cite{ah-ulcal-2021,blmn-mrp-2005,cgmz-hrdm-2016,bfn-cmsft-2019,Chan98,CHJ19,ghr-ond-2006,gkr-tpdr-2005,ls-focs-19}.}
\NotFullVer{%
The previous applications of tree covers were mostly limited to \emph{decomposable problems} where one can solve the problem by simply merging
the solutions for all the trees, for example like routing, spanners, nearest-neighbor searching, and network design~\cite{blmn-mrp-2005,bfn-cmsft-2019,CHJ19,gkr-tpdr-2005,ls-focs-19}.}
It seems challenging to apply a tree-cover-based approach to non-decomposable geometric problems such as the TSP or the EMWM, whose approximation algorithms are based on dynamic programming.
We create several new tools to overcome the challenges that arise in applying tree cover to EMWM.

When we work with a tree cover, portions of the min-net-cost augmenting path may appear at different levels in different trees.
As a result, maintaining $O(\mathrm{poly}\log n)$ sub-paths per cell will not be sufficient to faithfully reconstruct the min-net-cost augmenting path.
Nevertheless, we maintain only $O(\mathrm{poly}\log n)$ sub-paths per cell and settle with a weaker claim.
We define the \emph{$\theta$-adjusted cost} of an augmenting path $P$ to be its net cost
plus $\theta\cdot \norm{P}$, where $\norm{P}$ is the Euclidean arc length of $P$.
We show that using the sub-paths we can
compute an augmenting path $P$ whose $\theta$-adjusted cost is at most the $O(\theta \log n)$-adjusted cost of the min-net-cost augmenting
path (see \textsc{FindPath} procedure~(\ref{A:find-path}) and Lemma~\ref{L:findpath}).
Choosing such sub-optimal paths in tree covers leads to the following problems.

First, augmenting along a path $P$ now introduces an additive error $O(\theta\log n\cdot \norm{P})$ to the cost of the matching.
Large errors in the net-cost are introduced by non-matching edges on the endpoints of its sub-paths, as a sort of connection cost between sub-paths.
This error accumulates over the execution of the algorithm, thereby making it difficult to bound the total error in the matching cost.
We periodically repair the intermediate matching by canceling alternating cycles that have a negative $\theta$-adjusted cost.
Any such cycle $C$ has a net-cost at most $-\theta\norm{C}$; in other words, canceling this cycle reduces the matching cost by at least $\theta\norm{C}$.
We refer to them as \emph{reducing cycles}.
In fact, our algorithm only finds and cancels sufficiently many alternating cycles with a negative $\theta$-adjusted cost so that a weaker invariant is satisfied, i.e., the $(\theta\log n)$-adjusted costs of all cycles is non-negative.
By setting $\theta \coloneqq \eps/\log n$, we are able to bound the cost of the resulting matching to be $(1+\eps)M_\opt$.
We show that the time spent in canceling reducing cycles is bounded by the total time spent augmenting matchings.
All the earlier EMWM approximation algorithms computed a minimum-net-cost augmenting path under a suitable cost function and guaranteed that no negative net-cost cycles were created.

Second, to extract an augmenting path (or a reducing cycle) from a cell $\cell$ in the tree cover, the sub-paths are recursively expanded using those stored at the children cells of $\cell$.
A major complication from the overlapping grid cells of the tree-cover is that the augmenting path we compute --- by going down the hierarchy of cells and sub-paths --- may be self-intersecting, in which a matching edge appears multiple times.
Suppose $P = P_1 \circ P_2$, where $P_1$ and $P_2$ are expansions of sub-paths at the children cells of $\cell$.
Assuming $P_1$ and $P_2$ are simple, to guarantee that $P$ is also simple, we first check whether $P_1$ and $P_2$ share a matching edge. If the answer is yes, then $P$ contains a cycle $C$, which we can extract from $P$.
If~$C$ is a reducing cycle, then we update the current matching by ``canceling'' $C$ and ignoring $P$.
Otherwise we remove $C$ from $P$.  We repeat this step until $P$ becomes simple (or a reducing cycle is found).

Unfortunately, it is too expensive to first compute the self-inter\-secting path $P$ and then simplify it --- namely, check whether a matching edge in $P$ appears more than once, and if so, identify the cycle $C$ in~$P$ containing the edge and splice out $C$.
An important component of our algorithm is a dynamic data structure for quickly detecting cycles and simplifying augmenting paths.
Roughly speaking, we maintain intersection information about a collection of \emph{canonical paths} implicitly, using the hierarchy of residual graphs, each of which is guaranteed to be simple.
An augmenting path is constructed by concatenating these canonical paths at various levels.
At the heart of the data structure is a compact representation of canonical paths (despite the total complexity of canonical paths being quadratic, we are able to store them using near-linear space) and efficient procedures to detect whether two canonical paths share a matching edge (the intersection of two canonical paths may consist of many 
components), to extract cycles in the implicit representation of
an alternating path as concatenation of a sequence of canonical paths, and to splice a cycle from such a path.

The length-dependent penalty in the adjusted cost also acts  as a regularizer, and we use it to bound the total Euclidean length and total number of edges in all the augmenting paths and reducing cycles we compute.
The total running time remains near-linear (see \textsc{Augment} procedure~(\ref{A:augment})).

The paper is organized as follows: Section~\ref{S:overview} gives an overview of the overall algorithm; many of its details and analysis
are presented in Sections~\ref{S:ds_procedures} and~\ref{S:analysis}, respectively. The overall data structure based on tree-cover,
that finds an augmenting path and augments the matching is presented in Section~\ref{S:compression}, and its performance is analyzed in Section~\ref{S:ineq-proofs}.
Section~\ref{S:expanding} present the data structure for storing canonical paths compactly and the procedures that this data structure supports, including the intersection query.
\FullVer{Finally, Section~\ref{S:preprocess} describes the preprocessing step that preconditions the input.}
\NotFullVer{Due to space constraint, many proofs and details are omitted from the conference version.}

\section{Overall Algorithm}
\label{S:overview}



\paragraph{Preliminaries.}

Given a matching $M$, the \EMPH{residual graph} of $G$ with respect to $M$, denoted by $\GLOS[res-graph]{residual graph of $G$ with respect to $M$}{\vec{G}_M} = (V,E_M)$, is a directed graph on the vertices of $G$ in which all non-matching edges are directed from $A$ to $B$ and all matching edges from $B$ to $A$.
A vertex of $\vec{G}_M$ is called \EMPH{free} if it is not incident to any edge of $M$, and \EMPH{matched} otherwise.
%
An \EMPH{alternating path} (cycle) $\Pi$ in $G$ is a path whose edges alternate between non-matching and matching edges; $\Pi$ maps to a directed path (cycle) in $\vec{G}_M$.
%
Define the \EMPH{net cost} $\GLOS[net-cost]{net cost with respect to $\cost$ in $\vec{G}_M$}{\net\cost_M}\colon E_M \to \R$ on $\vec{G}_M$ as follows:
$\net\cost_M(a,b) \coloneqq \cost(a,b)$ if $(a,b) \not\in M$, and
$\net\cost_M(b,a) \coloneqq -\cost(a,b)$ if $(a,b) \in M$.
The net cost of any alternating path or cycle $\Pi$ is
\(
\net\cost_M(\Pi) = \sum_{e \in \Pi \setminus M} \cost(e) - \sum_{e \in \Pi \cap M} \cost(e).
\)
If $\Pi$ is an alternating path or cycle, we use the following shorthand for the arc length of $\Pi$ (as a polygonal curve):
\(
\EMPH{$\norm{\Pi}$} \coloneqq \sum_{e \in \Pi} \, \norm{e}.
\)
A simple (non-intersecting) alternating path $\Pi$ between two free
vertices is called an \EMPH{augmenting path}, and \EMPH{$M \oplus \Pi$} is a matching of size $\abs{M}+1$ from augmenting $M$ with $\Pi$.
If $\Pi$ is an alternating cycle, then $\abs{M \oplus \Pi} = \abs{M}$ instead.
The Hungarian algorithm repeatedly finds an augmenting path $\Pi$ in $\vec{G}_M$ of minimum net cost and augments the matching by $\Pi$.

\paragraph*{Adjusted cost.}
Let $c_0$ be a constant whose value will be chosen later.  Let~$M$ be any
fixed matching.
For a parameter $\theta \ge 0$, we define the \EMPH{$\theta$-adjusted cost} of an edge $e$ to be
\(
\GLOS[theta-adj-cost]{$\theta$-adjusted cost of edge $e$}{\adj_{\theta, M}(e)} \coloneqq \net\cost_M(e) + c_0\theta\cdot\cost(e),
\)
where $\net\cost_M(e)$ is the net cost of $e$ in the residual graph $\vec{G}_M$.
For a set ${X}$ of edges in $\vec{G}_M$, define
\(
\adj_{\theta, M}({X}) \coloneqq \sum_{e \in {X}} \adj_{\theta, M}(e).
\)
We can interpret $\adj_{\theta, M}$ as adding a \emph{regularizer} to the net cost of a residual path or cycle.
We fix two parameters: \EMPH{upper} $\GLOS[upper-e]{Upper $\e$}{\bar{\eps}} \coloneqq \frac{\eps}{c_1}$ and \EMPH{lower} $\GLOS[lower-e]{Lower $\e$}{\ubar{\eps}} \coloneqq \frac{\bar{\eps}}{c_2 \log n}$ where $c_1 \ge 8 c_0$ and $c_2 > 0$ are constants.
For a matching $M$, we define
\(
\GLOS[opt-adj-cost]{Minimum $\bar{\eps}$-adjusted cost among all augmenting paths with respect to $M$}{\adj^*_{\bar{\eps},M}} \coloneqq
	\min_{\Pi} \adj_{{\bar{\eps}}, M}(\Pi),
\)
where the minimum is taken over all augmenting paths with respect to $M$.
If the matching $M$ is clear from the context, we sometimes drop $M$ from
the subscript.
We call a cycle~$\Gamma$ in $\vec{G}_M$ \EMPH{reducing} if $\adj_{\ubar{\eps}}(\Gamma) < 0$.
We note that if $\adj_{\bar{\eps}}(\Gamma) < 0$, then $\Gamma$ is reducing (since $\ubar{\eps} \le \bar{\eps}$).
Intuitively, canceling a reducing cycle decreases the matching cost significantly relative to the cycle length (which is proportional to the amount of time required to cancel):
\(
    \adj_{\ubar{\eps}}(\Gamma) < 0
    \text{ implies }
    \net\cost(\Gamma) < -c_0\ubar{\eps}\cdot\norm{\Gamma}.
\)

\paragraph{Overview of the algorithm.}

We now present a high-level description of the algorithm.  We begin by performing a preprocessing step, the details of which can be found
\FullVer{in Section~\ref{S:preprocess}}
\NotFullVer{in the full version}
so that the input is ``well-conditioned'' at a slight increase in the cost of optimal matching (within an $(\eps/8)$-factor).
After this preprocessing step, which takes $O(n \log^2 n)$ time in total, we have point sets $A$ and $B$ that satisfy the following three properties:
\begin{enumerate}[label=P\arabic*., ref=P\arabic*]\itemsep=0pt
    \item \label{P:int_coords}
        All input points have integer coordinates.
    \item \label{P:monocolor}
        No integer grid point contains points of both $A$ and $B$.
    \item \label{P:opt_bounds}
        $\cost(M_\opt) \in {\bigl[ \frac{3\sqrt{d}n}{\eps}, \frac{9\sqrt{d}n}{\eps} \bigr]}$.
\end{enumerate}
Our goal is to compute an $\eps$-approximate matching of $A$ and $B$ satisfying (P1)--(P3) in $n \cdot (\eps^{-1}\log n)^{O(d)}$ time in the worst case.

After the preprocessing, the algorithm works in rounds, each of which increases the size of the matching by one.
The algorithm maintains the following \EMPH{cycle invariant} (cf.\ Corollary~\ref{C:cycleinv}):
\begin{enumerate}
[label={CI.}, ref=CI]
\item 
\label{I:cycle-invariant}
$\adj_{\bar{\eps}}(\Gamma) \ge 0$ for every alternating cycle $\Gamma$ at
the beginning of each round.
\end{enumerate}
Each round of the algorithm consists of two steps.
The first step computes an augmenting path $\Pi$ such that $\adj_{\ubar{\eps}}(\Pi) \le \adj^*_{\bar{\eps}}$ (recall that $\bar{\eps}=O(\ubar{\eps}\log n)$, so we only compute an approximate min-adjusted-cost augmenting path).
The second step updates $M$ by augmenting it by $\Pi$.
After augmentation, the matching may violate the cycle invariant~\ref{I:cycle-invariant}, so to reinstate the invariant the algorithm finds and cancels a sequence of simple reducing cycles $\Gamma_1, \ldots, \Gamma_k$ by updating $M \gets (((M \oplus \Gamma_1) \oplus \Gamma_2) \oplus \cdots)$.
To perform these steps efficiently, we design a data structure, described in 
Sections~\ref{S:compression} and~\ref{S:expanding}, that maintains $\vec{G}_M$ and supports the following two operations:
\begin{enumerate}[label=A\arabic*., ref=A\arabic*]\itemsep=0pt
\item \label{A:find-path}
$\textsc{Find-Path}()$: Returns an augmenting path $\Pi$ with $\adj_{\ubar{\eps}}(\Pi) \le \adj^*_{\bar{\eps}}$.
\item \label{A:augment}
$\textsc{Augment}(\Pi, M)$: Takes an augmenting path $\Pi$ and the current matching $M$ as input. First updates $M$ to $M \oplus \Pi$, then identifies and cancels a sequence of simple reducing cycles $\Gamma_1, \ldots, \Gamma_k$.
\end{enumerate}
%


The details of these operations are presented in Section~\ref{S:ds_procedures}. We prove the correctness of the algorithm and analyze its running time in Section~\ref{S:analysis}.
\section{Overall Data Structure}
\label{S:compression}

In this section we describe the overall data structure that maintains the current matching $M$ and residual graph $\vec{G}_M$, and that supports $\textsc{FindPath}$ and $\textsc{Augment}$.
The data structure constructs a hierarchical covering of $\reals^d$ (an instance of the \emph{tree cover} in $\reals^d$~\cite{gkr-tpdr-2005}) by overlapping hypercubes, called \emph{cells}, which is fixed and independent
of $M$.
For each cell $\cell$, it maintains a weighted, directed graph $G_\cell$ that depends on $M$ which can be viewed as a compressed representation of
the subgraph of $\vec{G}_M$ of size $\poly(\eps^{-1}\log n)$ induced by $(A \cup B) \cap \cell$.
The data structure detects negative cycles in $G_\cell$ and maintains shortest paths between pairs of nodes in $G_\cell$ if there are no negative cycles.
For each cell $\cell$, it also uses an auxiliary data structure that maps
a  negative cycle or ``augmenting path'' $\pi$ in $G_\cell$ to a simple (non-self-intersecting) negative cycle or augmenting path $\Pi$ such that $\smash{\adj_{\ubar{\eps}}(\Pi)}$ is at most the weight of $\pi$ in $G_\cell$; we refer to this map as an \emph{expansion} of $\pi$ in~$\vec{G}_M$.
Path $\Pi$ can be reported in time proportional to $\abs{\Pi}$.
	The data structure maintains a priority queue $\mathcal{Q}$ that stores the cheapest ``augmenting path'' of each $G_\cell$ and uses \textsc{Repair} procedure to update the information.

\mypara{Hierarchical covering and compressed graph.}
%
Let $\Grid_0$ be the $d$-dimensional integer grid, that is, the collection of hypercubes $[0, 1]^d + \ints^d$.
We build a hierarchy of ever coarser \EMPH{grids}  $\smash{\Grid_1, \dots, \Grid_{\log \Delta}}$.
Set $\GLOS[side-length]{Side-length of cells at level $i$}{\ell_i} \coloneqq 2^i$.
A hypercube in $\Grid_i$ has side-length $\ell_i$, and is obtained by merging $2^d$ smaller hypercubes of the grid $\Grid_{i-1}$ in the previous level.
For each $i$, we build a collection of \EMPH{cells} at level $i$ as follows:
For any hypercube $\cell \in \Grid_i$, we create $2^d$ cells at level $i$, namely, $\cell + \ell_{i-1} (b_1, \ldots, b_d)$ for all $d$-bits $b_1, \ldots, b_d \in \{0, 1\}$,
each corresponds to a shifting of~$\cell$ by either $+0$ or $+\ell_{i-1}$
in each dimension;
the cells at level $i$ is the union of $2^d$ different translations
$\Grid_i + \ell_{i-1} (b_1, \ldots, b_d)$ of $\Grid_i$.
%
Every cell at level $i$ has its boundary aligning with the grid boundaries of $\Grid_{i-1}, \Grid_{i-2}, \ldots, \Grid_0$.
As a consequence, cells are always perfectly tiled by lower-level grids, and cells at level $j$ are a refinement of 
those at level $i$ for all $j < i$.
The \EMPH{children} of $\cell$ at level $i$ are the $\smash{3^d}$ cells
at level $i-1$ contained in $\cell$.
We denote the set of children cells of $\cell$ as \EMPH{$\childr(\cell)$}.

Next, we describe the construction of compressed graphs.
We fix two parameters: \EMPH{height} $\GLOS[height]{Height of cell-covering dag}{h} \coloneqq c_3 \Ceil{\log n}$ and \EMPH{penalty} $\GLOS[penalty]{Penalty}{\pnlty} \coloneqq c_4 \ubar{\eps} = \Theta(\tfrac{\eps}{\log n})$, where $c_3, c_4 > 0$ are constants.
We translate the points slightly along each coordinate, say, by $\smash{\frac{\pnlty}{8\sqrt{d}}}$,
so that each point lies in the interior of a grid cell in $\Grid_i$ for any $\smash{i > \bigl\lceil \log \tfrac{\pnlty}{4\sqrt{d}} \bigr\rceil }$.
%
Let
\(
\GLOS[nonempty-cells]{Collection of nonempty cells at level $i$}{\Cover_i} \coloneqq \Set{\big. \text{level-$i$ cell $\cell$} \mid \cell \cap (A \cup B) \neq \varnothing }
\)
be the set of nonempty level-$i$ cells for each $i$; set $\GLOS[all-cells]{Collection of all nonempty cells}{\Cover} = \smash{\bigcup_{i=1}^{h} \Cover_i}$.
For each $\cell \in \Cover$ we construct a weighted directed bipartite graph $\GLOS[compressed-graph]{Compressed graph at cell $\cell$}{G_\cell} = (V_\cell, E_\cell)$ called the \EMPH{compressed graph}.

\mypara{Clustering and nodes of ${G_\cell}$.}
Each cell $\cell \in \Cover_i$ is partitioned into \EMPH{subcells} by the hypercubes of $\Grid_{i-\tau}$, where $\GLOS[offset]{Level offset between cells and subcells}{\tau} \coloneqq \bigl\lceil \log_2(4\sqrt{d}/\pnlty) \bigr\rceil$ is the \EMPH{level-offset}.\footnote{In this paper, the base of $\log$ is always $2$.}
The diameter of each subcell is at most $\smash{\frac{\pnlty}{4}\ell_i}$.
%
For each subcell $\subcell$ of $\cell$, let $\GLOS[A-subcell]{Points of $A$ in subcell $\subcell$}{A_\subcell} \coloneqq A \cap \subcell$ and $\GLOS[B-subcell]{Points of $B$ in subcell $\subcell$}{B_\subcell} \coloneqq B \cap \subcell$.
We refer to $A_\subcell, B_\subcell$ as the \EMPH{$A$-clusters} and \EMPH{$B$-clusters} respectively.
We call a cluster $A_\subcell$ or $B_\subcell$ \EMPH{unsaturated} if at least one of its points is free; otherwise we call it \EMPH{saturated}.
Set $\GLOS[A-collection]{Collection of nonempty $A_\subcell$s in cell $\cell$}{\mathcal{A}_\cell}$ and $\GLOS[B-collection]{Collection of nonempty $B_\subcell$s in cell $\cell$}{\mathcal{B}_\cell}$ to be the collections of nonempty $A$- and $B$-clusters of $\cell$, respectively.
%
The set of nodes of $G_\cell$ is $\GLOS[compressed-graph-vertices]{Vertices of compressed graph
$G_\cell$}{V_\cell} \coloneqq \mathcal{A}_\cell\cup\mathcal{B}_\cell$; $\EMPH{s} \coloneqq \abs{V_\cell} = (\eps^{-1} \log n)^{O(d)}$.
We note that each cell $\cell \in \Cover_0$ either contains only points of $A$
or points of $B$.
For level $i \ge 1$, let $\subcell$ be a subcell of $\cell$.
Then
\(
A_\subcell = \bigcup_{\Delta \in \childr(\subcell)} A_{\Delta}
\)
and
\(
B_\subcell = \bigcup_{\Delta \in \childr(\subcell)} B_{\Delta},
\)
where  $\GLOS[children-subcells]{Children subcells of $\subcell$}{\childr(\subcell)}$ are the \EMPH{children subcells} of $\subcell$.
Hence, a cluster of~$\cell$ is obtained by merging the at most \smash{$2^d$} clusters of children of $\cell$ that contain $\subcell$.

\mypara{Arcs, compressed paths, and expansions.}
Graph $G_\cell$ is a directed complete bipartite graph with arc set $\GLOS[compressed-graph-arcs]{Arcs of compressed graph $G_\cell$}{E_\cell} \coloneqq \mathcal{A}_\cell\times\mathcal{B}_\cell \cup \mathcal{B}_\cell\times\mathcal{A}_\cell$.
%
There are two types of arcs in $E_\cell$:
an arc between two clusters $A_{\subcell}$ and $B_{\subcell'}$ is a \EMPH{bridge arc} if no child cell of $\cell$ contains both of $\subcell$ and $\subcell'$;
otherwise it is an \EMPH{internal arc}.

\begin{figure}[htb]
	\centering
    \includegraphics[width=0.25\textwidth, page=1]{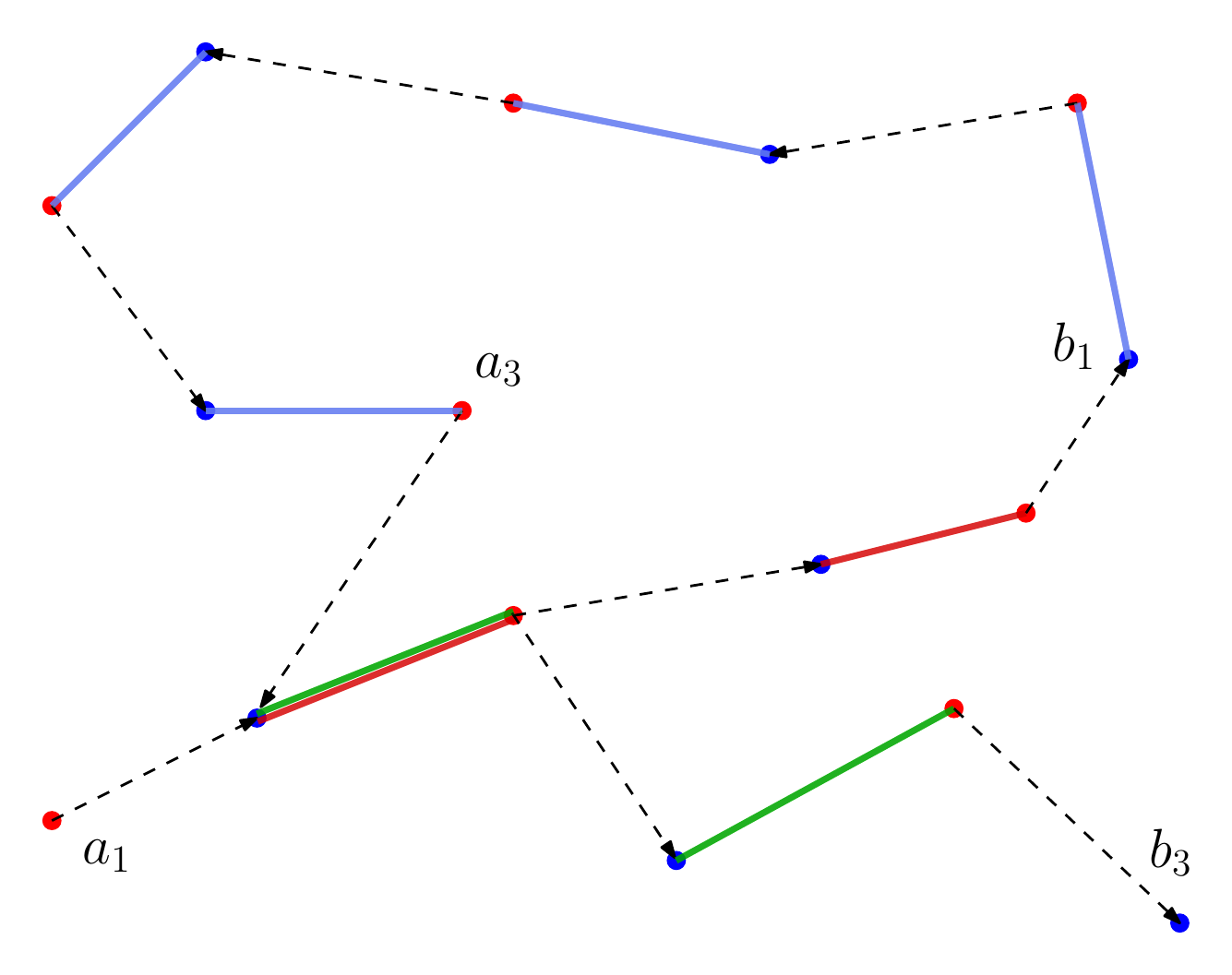}
    \caption{
	    Concatenation of three expansions $\Phi_1$ (red), $\Phi_2$ (blue), $\Phi_3$ (green); 
	    solid edges are matching edges.
	    The ending (starting) tip of $\Phi_1$ ($\Phi_3$) is changed to connect it to $\Phi_2$.
	    $\Phi_1$ and $\Phi_3$ share a matching edge.}
    \label{F:FTE-concat}
\end{figure}

We will soon define a recursive weight function $w_\cell$ on the arcs in $E_\cell$.
Using this weight function, we compute a minimum-weight path $\GLOS[min-weight-path]{Minimum weight path between $X$ and $Y$ in graph $G_\cell$}{\pi_\cell(X, Y)}$ between all pairs of nodes/clusters $(X, Y)$ in $G_\cell$; we refer to $\pi_\cell(X, Y)$ as a \EMPH{compressed path}.
For every pair $(X,Y) \in E_\cell$, we also compute a simple path in $\vec{G}_M$ 
that begins at a point in $X$ and ends at a point in $Y$, which we call the \EMPH{expansion} of $\pi_\cell(X,Y)$, denoted by $\GLOS[expansion]{Expansion of $\pi_\cell(X,Y)$, a simple alternating path from $X$ to $Y$}{\Phi_\cell(X,Y)}$.
We refer to the endpoints of $\Phi_\cell(X,Y)$ as \EMPH{tips}. If $(X,Y)=(B_\subcell,A_{\subcell'})$, then the first and last edges  of $\Phi_\cell(B_\subcell,A_{\subcell'})$ are matching edges and the tips are never changed.
On the other hand,  if $(X,Y)=(A_\subcell,B_{\subcell'})$ then the first and last edges of $\Phi_\cell(B_\subcell,A_{\subcell'})$ are non-matching edges and we may change the tips as needed.
That is, if the first (resp.\ last) edge of $\Phi_\cell(A_\subcell,B_{\subcell'})$ is $(a, b)$ and 
$a'\in A_\subcell$ (resp.\ $b'\in B_{\subcell'})$ is another point,
then we may replace $(a,b)$ with $(a',b)$ (resp. $(a,b')$).

The asymmetry in the definition of $\Phi_\cell(B_\subcell, A_{\subcell'})$ and $\Phi_\cell(A_\subcell, B_{\subcell'})$ 
allows us to concatenate the expansion of two shortest paths in $G_\cell$ that share a node.
Suppose $\Phi_1 = \Phi_\cell(B_{\subcell_1}, A_{\subcell_2})$,
$\Phi_2 = \Phi_\cell(A_{\subcell_2}, B_{\subcell_3})$,
$b_1$ and $a_1$ the starting and ending tips of $\Phi_1$ respectively,
then we change the starting tip of $\Phi_2$ to $a_1$ and $b_3$.  
By concatenating this modified $\Phi_2$ with $\Phi_1$, we obtain an alternating path in $\vec{G}_M$ from the starting tip of $\Phi_1$ to 
the ending tip of $\Phi_2$ (see Figure~\ref{F:FTE-concat}).
For brevity, we will simply write this concatenation as $\Phi_1 \circ \Phi_2$ in the rest of the text. 
If $\subcell_3=\subcell_1$, i.e., $\pi_\cell(B_{\subcell_1},A_{\subcell_2})\circ\pi_\cell(A_{\subcell_2}, B_{\subcell_3})$ is a cycle in $G_\cell$, then by \emph{cyclically concatenating} $\Phi_1$ and $\Phi_2$, i.e, changing the starting tip of $\Phi_2$ to $a_1$ and the ending tip of $\Phi_2$ to $b_1$, $\Phi_1\circ\Phi_2$ is an alternating cycle in $\vec{G}_M$.

There is a mutual recursion between the definition of arc weights and the construction of expansions: arc weights are defined using the residual cost of lower-level expansions, and expansions are constructed based on shortest paths $\pi_\cell(X,Y)$, described below.

\paragraph{Stability.}
During both the shortest path computation on $G_\cell$ and the construction of expansions we may discover a reducing cycle $\Gamma$ in \smash{$\vec{G}_M$}, in which case we update $M$ by canceling $\Gamma$ (i.e., setting $M \from M\oplus \Gamma$) and rebuilding the data structure at $\cell$ and its descendants (see \textsc{Repair} in Section~\ref{S:ds_procedures}).
We say cell $\cell$ is \EMPH{stable} if
(i) the children of $\cell$ are stable, and
(ii) no reducing cycle was detected while computing $\pi_\cell(\cdot, \cdot)$ and $\Phi_\cell(\cdot, \cdot)$ for all pairs of nodes in $G_\cell$.

\paragraph{Arc weights.}

Assuming all children cells of $\cell$ are stable, we define \EMPH{arc weights} $\GLOS[edge-weight]{Arc weights on $G_\cell$}{w_\cell}\colon E_\cell \to \R$ recursively as follows:
\begin{itemize}\itemsep=0pt
    \item \textbf{Bridge arcs}:
	    Let $(X,Y)$ be a bridge arc. Let $\cnter_X$ (resp.\ $\cnter_Y$) be the center of the subcell 
		containing the cluster $X$ (resp.\ $Y$).  For $(X,Y)=(A_\subcell,B_{\subcell'})$, we set
        \begin{equation} \label{E:weight_nonmatch_bridge}
            w_\cell(X, Y) \coloneqq \norm{\cnter_X - \cnter_Y} + \pnlty\ell_i
        \end{equation}
        if $X\times Y \setminus M\ne\varnothing$ and $\infty$ otherwise. For $(X,Y)=(B_\subcell,A_{\subcell'})$, 
        \begin{equation} \label{E:weight_match_bridge}
            w_\cell(X, Y) \coloneqq -\norm{\cnter_X - \cnter_Y} + \pnlty\ell_i 
        \end{equation}
if $(X \times Y) \cap M \neq \varnothing$ and $\infty$ otherwise.

    \item \textbf{Internal arcs}:
        If a child $\cell'$ of $\cell$ contains both $\subcell$ and $\subcell'$, then we set
        \begin{equation}
		\label{E:weight_match_internal}
		    w_\cell(X_\subcell, Y_{\subcell'}) \coloneqq \min_{\substack{\Delta \in\childr(\subcell), \Delta' \in \childr(\subcell') \\ \cell' \in \childr(\cell): \subcell, \subcell' \subset \cell'}} \adj_{\ubar{\eps}}\Paren{\Phi_{\cell'}(X_\Delta, Y_{\Delta'})} + \pnlty\ell_i.
		\end{equation}
		We call the triple $(\cell',\Delta,\Delta')$ that realizes the minimum in (\ref{E:weight_match_internal})
		as the \EMPH{weight-certificate} of the internal arc $(X_\subcell, Y_{\subcell'})$.
\end{itemize}

\paragraph{Expansions.}

We now define the expansions $\Phi_\cell(X,Y)$ using the weights of arcs in $G_\cell$ and their weight certificates.
We first define an expansion \EMPH{$\Phi(\eta)$} of each arc $\eta\in E_\cell$.
If $\eta$ is a bridge arc of the form $\eta = (A_\subcell,B_{\subcell'})$, then we choose $\Phi(\eta)$ to be an arbitrary pair $(a,b)\in A_\subcell\times B_{\subcell'}$ with the caveat that if $A_\subcell$ or $B_{\subcell'}$ has a free point then we choose $a$ or $b$ to be a free point;%
\FullVer{%
and if it is a bridge arc of the form $\eta=(B_\subcell,A_{\subcell'})$ then we choose $\Phi(\eta) = \argmax_{e \in M \cap (B_\subcell \times A_{\subcell'})} \norm{e}$ (the longest matching edge in the cluster).}
\NotFullVer{%
and if it is a bridge arc of the form $\eta=(B_\subcell,A_{\subcell'})$ then we choose $\Phi(\eta) = \argmax_{e \in M \cap (B_\subcell \times A_{\subcell'})} \norm{e}$.}
Finally, if $\eta$ is an internal arc with $(\cell',\Delta,\Delta')$ as the weight certificate of $\eta$, then we set $\Phi(\eta) \coloneqq \Phi_{\cell'} (X_\Delta, Y_{\Delta'})$.

For a given pair $(X,Y) \in E_\cell$, we compute $\Phi_\cell(X,Y)$ as follows: Suppose $\pi_\cell(X,Y)=\Seq{\eta_1, \ldots,\eta_k}$, where each $\eta_i$ is an arc of $G_\cell$.  Let 
$\tilde\Phi \leftarrow  \Phi(\eta_1)\circ\cdots\circ\Phi(\eta_k)$, where concatenation is as defined above. 
Albeit each $\Phi(\eta_i)$ being a simple path, $\tilde\Phi$ may not be simple, i.e., some matching edges may appear more than once, so we cannot simply set $\Phi_\cell(X,Y)$ to $\tilde\Phi$. Instead, we use a greedy algorithm to simplify 
$\tilde\Phi$ by repeatedly removing cycles, as follows: We check whether any matching edge in $\tilde\Phi$ appears
more than once. If there is no such edge, then we stop and set $\Phi_\cell (X,Y) \from \tilde\Phi$. 
Otherwise, suppose $\tilde\Phi = \Pi_1 \circ \seq{e} \circ \Pi_2 \circ \seq{e} \circ\Pi_3$, where $e$ is a matching edge.
Then $C \coloneqq \Seq{e} \circ \Pi_2$ is an alternating cycle in $\vec{G}_M$ (see Figure~\ref{F:FTE-concat}). 
If $\adj_{\ubar{\eps}}(C) <0$, i.e., $C$ is a reducing cycle, then we compute a simple reducing subcycle of $C$ using the procedure \textsc{SimpleReducingSubcycle} described below.
Otherwise we set $\tilde\Phi \leftarrow \Pi_1 \circ \seq{e} \circ \Pi_3$, and repeat the above step.
The correctness of the overall algorithm does not depend on the order in which we find the duplicate matching edges. We refer to this procedure of computing $\Phi_\cell(X,Y)$ from $\pi_\cell(X,Y)$ as \textsc{ConstructExpansion}.
Maintaining expansions in a data structure so that they can be computed efficiently is a major component of our algorithm and is described in detail in Section~\ref{S:expanding}.

The following lemmas summarize
the relationship between the adjusted cost of paths in $\vec{G}_M$ and the weights of paths in $G_\cell$.

\begin{lemma}[Expansion Inequality]
\label{L:expansion}
	Let $\cell$ be a stable cell, let $(X,Y) \in E_\cell$.
Then
  $\adj_{\ubar{\eps}}(\Phi_{\cell}(X, Y)) \le w_\cell(\pi_{\cell}(X,Y))$. 
  For $(X,Y) = (A_\subcell,B_{\subcell'})$, the inequality holds for all choices of its tips in $A_\subcell \times B_{\subcell'}$.
\end{lemma}


\begin{corollary}
\label{C:negativecycle_reducingcycle}
Suppose all children of $\cell$ are stable.
Let $C \coloneqq \Seq{\eta_1, \ldots, \eta_k} \subseteq E_\cell$
be a negative cycle in $G_\cell$, and let $\Gamma \coloneqq \Phi(\eta_1) \circ \cdots \circ \Phi(\eta_k)$ be the 
	alternating cycle formed by cyclically concatenating the expansions of $\eta_1, \ldots, \eta_k$.
Then, $\adj_{\ubar{\eps}}(\Gamma) < 0$.
\end{corollary}

\begin{lemma}
\label{L:container_exists}
Let $\Pi$ be either a reducing cycle, or the augmenting path (in $\vec{G}_M$) with the minimum $\bar{\eps}$-adjusted cost $\adj^*$.
Then there exists a level $i \in \set{0,\ldots,h}$ and $\cell \in \Cover_i$ such that $\Pi$ lies completely in $\cell$.
\end{lemma}

\begin{lemma}[Lifting Inequality]
\label{L:lift}
Let $\Pi$ be an alternating path in $\vec{G}_M$ from point $p$ to point $q$ (possibly $p=q$ in the case $\Pi$ is a cycle) that is completely contained in a cell of level $\le h$.
Let $\cell$ be a cell at the smallest level that contains $\Pi$ and
$X$ (resp.\ $Y$) the cluster in $V_\cell$ containing $p$ (resp.\ $q$).
Then either $\cell$ is not stable or
\(
	w_\cell(\pi_\cell(X, Y)) \le \adj_{\bar{\e}}(\Pi).
\)
\end{lemma}

\mypara{Putting everything together.}
In view of the above lemmas, the data structure works as follows:
If $\cell$ is not stable (but all its children are), then the data structure returns a reducing cycle $\Gamma_\cell$ in $\vec{G}_M$
that acts as a witness of its instability.
If $\cell$ is stable, then for every pair $(X,Y) \in E_\cell$, the data structure maintains a 
minimum-weight path $\pi_\cell(X,Y)$ and its expansion $\Phi_\cell(X,Y)$.
If both $A_\subcell, B_{\subcell'}$ are unsaturated, then we call $\pi_\cell(A_\subcell, B_{\subcell'})$ an \EMPH{augmenting path} in $G_\cell$.
Given an augmenting path $\pi_\cell(A_\subcell, B_{\subcell'})$ in $G_\cell$, our choice of the expansion of a bridge arc ensures that $\Phi_\cell(A_\subcell, B_{\subcell'})$ is an augmenting path in $\vec{G}_M$.
Let
\(
    \GLOS[zeta]{Subcell pair that realizes minimum $\ubar{\e}$-adjusted cost of augmenting path $\Pi_{\subcell, \subcell'}$ in residual graph $\vec{G}_M$ among all pairs of unsaturated $A_\subcell$ and $B_{\subcell'}$}{(\subcell_\cell, \subcell_\cell')}
        \coloneqq \argmin_{\subcell, \subcell'} \adj_{\ubar{\eps}}(\Phi_\cell(A_\subcell, B_{\subcell'}))
\)
where the minimum is taken over all cluster pairs $A_\subcell, B_{\subcell'}$ of $\cell$ such that both are unsaturated.
If $G_\cell$ does not have any augmenting path, the pair $(\subcell_\cell, \subcell_\cell')$ is undefined.
Set $\GLOS[all-candidate-pairs]{Set of $(\subcell_\cell, \subcell_\cell')$ over all pairs $A_\subcell, B_{\subcell'}$ of $\cell$}{\Pairs} \coloneqq \left\{ (\subcell_\cell, \subcell_\cell') \mid \cell \in \Cover \right\}$.
We store $\Pairs$ in a priority queue with $\adj_{\ubar{\eps}}(\Phi_\cell(\subcell_\cell, \subcell_\cell'))$ as the key of $(\subcell_\cell, \subcell_\cell')$.

\textsc{FindPath} and \textsc{Augment} procedures are described in Section~\ref{S:ds_procedures}.
The information stored at $\cell$ depends on $M \cap ((A\cap\cell) \times (B\cap\cell))$.
So whenever the matching edges change (e.g. by \textsc{Augment}), we update the information stored at the corresponding $\cell$.
The $\textsc{Repair}(\cell)$ procedure, also described in Section~\ref{S:ds_procedures}, updates the data structure at $\cell$.
Initially $M = \varnothing$, and the data structure can be built by calling $\textsc{Repair}$ at all cells of $\Cover$ in a bottom-up manner.

\section{Maintaining Expansions of Compressed Paths}
\label{S:expanding}

We now describe the algorithms and data structure for computing and maintaining the expansions of compressed paths at all cells.
We begin by describing a high-level representation of expansions and the two high-level procedures 
\textsc{ConstructExpansion} and \textsc{SimpleReducingSubcycle} needed for computing them.
Next, we describe the data structure to maintain the expansions compactly.
Finally, we describe the 
\textsc{Intersect}, \textsc{Report}, and \textsc{AdjCost} procedures as well as 
the operations on the data structure needed by the high-level procedures.

Recall that cells in $\Cover$ are processed in a bottom up manner.
We focus on computing expansions at a cell $\cell$, assuming (i) they have been computed at all children cells of 
$\cell$, (ii) the children of $\cell$  are stable, (iii)
compressed paths between all pairs of subcells of $\cell$ have been computed, and (iv) no negative cycles have been detected in $G_\cell$.
For an internal arc $\gamma = (A_\subcell, B_{\subcell'})$ at $\cell$, the expansion of $\gamma$, $\Phi(\gamma)$, 
is $\Phi_{\cell'}(A_\Delta, B_{\Delta'})$ for some child cell $\cell'$ of $\cell$ and 
children clusters $A_\Delta, B_{\Delta'}$ of $A_\subcell, B_{\subcell'}$, where $(\cell',\Delta,\Delta')$ is the weight certificate of $\gamma$.
We have $\Phi(\gamma)$ at our disposal when we compute expansions at $\cell$.

\subsection{Expansions and pathlets}
\label{SS:pathlet}

\mypara{Representation of expansions.}
Let $\Seq{b_1, a_1, \ldots, b_k, a_{k}}$ be a path in $\vec{G}_M$ from $b_1 \in B$ to $a_k \in A$, or a cycle (with the 
edge $(a_k,b_1)$ also being present).
The path (cycle) can be specified by the sequence $\Seq{(b_1, a_1), \ldots, (b_{k}, a_{k})}$ of matching edges.
Conversely, since $(a, b) \in \vec{G}_M$ if $(b, a) \not\in M$ for any pair $(a, b) \in A \times B$, any sequence $\Phi \coloneqq \Seq{e_1, e_2, \ldots, e_k}$ of matching edges defines a path
\(
\Seq{b_1, a_1, \ldots, b_k, a_k}
\)
in~$\vec{G}_M$, where $e_i = (b_i, a_i)$.
$\Phi$ also defines a unique cycle in $\vec{G}_M$.
If the first or the last edge of an alternating path is a non-matching edge, then the path can be represented by its sequences of matching edges and its tips. For example, an alternating path $\Seq{a_0, b_1, a_1, \ldots, b_k, a_{k}, b_{k+1}}$ can 
be represented by the edge sequence $\Seq{(b_1, a_1), \ldots, (b_{k}, a_{k})}$ and the starting and ending tips 
$a_0, b_{k+1}$, respectively.
We represent an alternating paths/cycle $\Pi$ as a tuple called the \EMPH{compact representation}, denoted by $\langle\Pi\rangle$, comprised of:
\begin{itemize}\itemsep=0pt
    \item a sequence $\Seq{e_1, e_2, \ldots, e_k}$ of edges in $M$, which we refer to as the \EMPH{matching edge sequence} (MES);
        this sequence is empty if $\Pi$ consists of a single non-matching edge; 
    \item a \emph{flag} which is set to $1$ if $\Pi$ is a cycle; and
    \item the \emph{tips} if the first and/or last edge of $\Pi$ is non-matching (otherwise \textit{null});
        tips are always \textit{null} if $\pi$ is a cycle.
\end{itemize}

\mypara{Pathlets.}
For any MES $\Phi = \Seq{e_1, \ldots, e_k}$ and two indices $i \le j$, we define the \EMPH{splice} operators
\(
\GLOS[incl-splice]{Inclusive splice operators}{e_i \spL \Phi \spR e_j} \coloneqq \langle e_i, e_{i+1}, \ldots, e_{j-1}, e_j \rangle
\)
and
\(
\GLOS[excl-splice]{Exclusive splice operators}{e_i \spoL \Phi \spoR e_j} \coloneqq \langle e_{i+1}, e_{i+2}, \ldots, e_{j-2}, e_{j-1} \rangle.
\)
We mix the inclusive and exclusive splices when convenient, e.g., $e_i \spoL \Phi \spR e_j = \langle e_{i+1}, \ldots, e_{j-1}, e_j \rangle$.
We call any contiguous subsequence of $\Phi$ a \EMPH{pathlet} of $\Phi$.
Obviously, $\Phi$ is a pathlet of itself.
If $\phi$ is a pathlet of~$\Phi$, we also allow further restriction of $\phi$ using splicing, e.g., to create new pathlets $\phi' \subseteq \phi$ of $\Phi$.
We say that $\phi$ \EMPH{originates} from $\Phi$, 
and we use ``pathlet of an expansion'' as shorthand for ``pathlet of the MES of an expansion.''
The pathlet of an expansion is always simple, since an expansion's MES is simple.
Later, we compose new MES by concatenating multiple (individually) simple pathlets --- these concatenated sequences need not be simple.
Two pathlets \emph{intersect} if they have a common matching edge.

Although the compact representation and pathlets can be applied to any path/cycle in $\vec{G}_M$, we will deal mainly with those of three types of paths: (i) an expansion, (ii) pathlets of an expansion, and (iii) concatenations of pathlets of expansions.
We work mostly with the MES except when we need to compute the adjusted cost (i.e., tips matter), so with a slight abuse of notation we will not distinguish between an expansion and its MES and use $\Phi$ to denote both.
Note that (i) the MES of $\Phi_1\circ\Phi_2$ is the concatenation of MES's of $\Phi_1, \Phi_2$, and (ii) an alternating path is non-simple if and only if its MES contains duplicate (matching) edges.
These two simple observations will be crucial for our data structure.

\mypara{Operations on pathlets.}
MES's and pathlets are maintained using a data structure, which consists of a collection of trees, called MES-trees and pathlet-trees, respectively, and Boolean look-up tables,
described  below in Section~\ref{SS:pathlet_compact}. 
This data structure supports the following operations. Here we assume that a pathlet is represented compactly using $O(\log n)$ size (see  Section~\ref{SS:pathlet_compact}).

\begin{itemize}\itemsep=0pt
    \item $\textsc{Intersects}(\phi_1, \phi_2)$:
        Given two pathlets $\phi_1$ and $\phi_2$ that originate from expansions of descendant cells of $\cell$, report whether $\phi_1 \cap \phi_2 \neq \varnothing$.

    \item $\textsc{LastCommonEdge}(\phi_1, \phi_2)$:
        Given two pathlets $\phi_1$ and $\phi_2$ that originate from lower-level expansions where $\phi_1 \cap \phi_2 \neq \varnothing$, return \emph{references} to four edges\footnotemark:
        $e_1$, the last edge of $\phi_1$ in the common intersection;
        $e_2$, the copy of $e_1$ in $\phi_2$;
        $e_3$, the predecessor of $e_1$ in $\phi_1$ (maybe \textit{null});
		$e_4$, the predecessor of $e_2$ in $\phi_2$ (maybe \textit{null}).

        \footnotetext{
        We will explain the specific form of the references to these edges when we describe the data structure.
        We use the references to $e_1, e_2$ (resp.\ $e_3, e_4$) to implement inclusive (resp.\ exclusive) splices involving $e_1$ in both $\phi_1$ and $\phi_2$. None of the procedures will be invoking left exclusive splice operation, i.e., of the form $e\spoL\phi$, so we do not need references to the successor edges.
        }

    \item $\textsc{Median}(\phi)$:
        Given a pathlet $\phi$, return a reference to the median edge of $\phi$.
        That is, given $\phi \coloneqq \langle e_1, e_2, \ldots, e_k \rangle$, return a reference to $e_{\Ceil{k/2}}$.

    \item $\textsc{AdjCost}(\langle\Pi\rangle)$:
        Given a possibly non-simple path $\Pi \in \vec{G}_M$ in its compact representation, where its MES is the concatenation of up to $s$ pathlets,
		return $\adj_{\ubar{\eps}}(\Pi)$.


        
\item $\textsc{Report} (\langle\Pi\rangle)$
        Given a simple path/cycle $\Pi \in \vec{G}_M$, that is the concatenation of up to $s$ pathlets of expansions $\phi_1, \dots, \phi_k$, return the sequence of edges in $\Pi$.

\item \textsc{Splice}$(\phi, e, \bowtie)$:
	Given a pathlet $\phi$, an edge $e$ of $\phi$, and $\bowtie\,\, \in \set{\spL,\spR,\spoR}$, splice the pathlet $\phi$ at $e$.

\item \textsc{Concatenate}$(\phi_1, \phi_2, \ldots, \phi_t)$:
	Construct an MES composed of $\phi_1\circ\phi_2\circ\cdots\circ\phi_t$.
\end{itemize}

Let $t(n)$ denote the maximum time taken by the above procedures except \textsc{Report}. We will see below in 
Section~\ref{SS:pathlet_ops} that $t(n)=(\eps^{-1}\log n)^{O(d)}$ and \textsc{Report} takes $O(t(n)+kh)$ 
time where $k$ is the length of the path returned by the procedure.

\subsection{High-level simplification procedures}
\label{SS:high_level}

We now describe the two main  procedures $\textsc{ConstructExpansion}$ and $\textsc{SimpleReducingSubcycle}$.

\mypara{\textsc{ConstructExpansion}.}
Let $\pi_\cell(X,Y) = \langle \eta_1, \ldots, \eta_k \rangle$ be a shortest path in $G_\cell$,
where each $\eta_i$ is an arc of $G_\cell$. We compute the MES of $\Phi_\cell(X,Y)$ and set 
the starting (resp.\ ending) tip of $\Phi_\cell(X,Y)$ to that of $\Phi(\eta_1)$ (resp.\ $\Phi(\eta_k))$.
Let $\phi_i$ be the MES of $\Phi(\eta_i)$, for $1\le i\le k$. We initially set
\begin{equation}
\label{E:pathlet_ie}
\tilde\Phi 
    \from \phi_1 \circ \cdots \circ \phi_k .
\end{equation}
We process the $\phi_i$'s in sequence and grow a pathlet sequence 
$\Xi = \langle \phi'_1, \ldots, \phi'_t \rangle$ where $\phi'_1 \circ \cdots \circ \phi'_t$ is simple.

Initially, $\Xi = \varnothing$ and the first pathlet we process is $\phi_1$.
To process $\phi_i$, we compare it against each $\phi'_j \in \Xi$ in ascending order to determine whether $\phi'_0 \circ \cdots \circ \phi'_t \circ \phi_i$ is simple.
Specifically, we query $\textsc{Intersects}(\phi_i, \phi'_j)$ to find the first pathlet $\phi'_j \in \Xi$ that shares a matching edge with $\phi_i$.
If there is no intersection with any pathlet in $\Xi$, we simply set $\phi'_{t+1} \from \phi_i$, append $\phi'_{t+1}$ to $\Xi$, and continue on to $\phi_{i+1}$.
If there is an intersection against $\phi'_j$, we find $e_1$, the \emph{last} edge in $\phi_i$ that intersects $\phi'_j$, by invoking $\textsc{LastCommonEdge}(\phi_i, \phi'_j)$.
Then, $\phi'_1 \circ \cdots \circ \phi'_{j-1} \circ (\phi'_j \spoR e_1) \circ (e_1 \spL \phi_i)$ is a simple sequence of matching edges and
\[
C \coloneqq (e_1 \spL \phi'_j) \circ \phi'_{j+1} \circ \cdots \circ \phi'_t \circ (\phi_i \spoR e_1)
\]
is a cycle.\footnotemark
\footnotetext{
Note that the exclusive splices (e.g., \smash{$\phi'_j \spoR e_1$}) may be empty pathlets.
If that is the case, we simply drop the empty pathlet from the concatenation sequence.
}
Next, compute $\adj_{\ubar{\eps}}(C)$ using $\textsc{AdjCost}(\langle C\rangle)$.
If $C$ is reducing, we abort $\textsc{ConstructExpansion}$ and return the simple reducing cycle $\smash{\hat{C}}$ by calling $\textsc{SimpleReducingSubcycle}(C)$.
If $C$ is not reducing, then we update $\phi'_j = (\phi'_j \spoR e_1)$, $\phi'_{j+1} = (e_1 \spL \phi_i)$, and set $\Xi \from \langle \phi'_1, \ldots, \phi'_j, \phi'_{j+1} \rangle$.
Then, we continue on to $\phi_{i+1}$.
If all elements of $\tilde\Phi$ are processed without aborting and $\Xi = \langle \phi'_0, \ldots, \phi'_t \rangle$, then we return the simple sequence $\phi'_1 \circ \phi'_2 \circ \cdots \circ \phi'_t$ as the MES of $\Phi_\cell(X,Y)$.

Here we assume that each input pathlet $\phi_i$ is represented as a pathlet tree, intermediate pathlets are also maintained using
pathlet trees, pathlets are spliced using \textsc{Splice}, and  \textsc{Concatenate}$(\phi'_1, \ldots, \phi'_t)$ is called at the end to represent the MES as an MES tree.

\mypara{\textsc{SimpleReducingSubcycle}.}
This procedure is either called by $\textsc{ConstructMES}$ or by $\textsc{Repair}$ to expand a negative compressed cycle $\varphi$ in $G_\cell$.
In the latter case, we first construct an intermediate expansion of $\varphi$ of the form (\ref{E:pathlet_ie}) as above, so assume that (MES of) the reducing cycle is represented as a pathlet sequence $C = \Seq{\phi_1, \ldots, \phi_k}$.
Like $\textsc{ConstructExpansion}$, the general case of this procedure processes the pathlets of the input in sequence while building a simple prefix.
Given a reducing cycle (as a pathlet sequence) $C$, we grow a pathlet sequence $\Xi = \langle \phi'_1, \ldots, \phi'_t \rangle$ where $\phi'_1 \circ \cdots \circ \phi'_t$ is simple.
The case of $k\le 2$ is a base case, handled differently, so assume first that $k \ge 3$.

Initially $\Xi = \varnothing$, and the first element we process is $\phi_1$.
Suppose we are processing $\phi_i$ in $C$.
For each $\smash{\phi'_j} \in \Xi$ in reverse order, we query $\smash{\textsc{Intersects}(\phi_i, \phi'_j)}$.
If we fail to find any intersections between $\Xi$ and $\phi_i$, then $\phi'_1 \circ \cdots \phi'_t \circ \phi_i$ is simple, so we set $\phi'_{t+1} \from \phi_i$, append $\phi'_{t+1}$ to $\Xi$, and continue on to $\phi_{i+1}$.
Otherwise, let $\phi'_j$ be the last element of $\Xi$ to intersect $\phi_i$, and we invoke $\textsc{LastCommonEdge}(\phi_i, \phi'_j)$ to acquire $e_1$, the \emph{last} edge of $\phi_i$ in the intersection.
There are two subcycles about $e_1$:
\[
\begin{aligned}
C_0 &\coloneqq (e_1 \spL \phi'_j) \circ \phi'_{j+1} \circ \cdots \circ \phi'_{i-1} \circ (\phi_i \spoR e_1) \\
C_I &\coloneqq \phi'_1 \circ \cdots \circ (\phi'_j \spoR e_1) \circ (e_1 \spL \phi_i) \circ \phi_{i+1} \circ \cdots \circ \phi_t
\end{aligned}
\]
Neither $C_0$ nor $C_I$ need be simple, but:
\begin{itemize}\itemsep=0pt
    \item In $C_0$, the only potentially intersecting pathlets are $e_1 \spL \phi'_j$ and $\phi_i \spoR e_1$.
    \item In $C_I$, $(\phi'_j \spoR e_1) \cap (e_1 \spL \phi_i) = \varnothing$, so the number of pairs of intersecting component pathlets in $C_I$ is strictly less than the number in $C$.
\end{itemize}
Since $C$ was reducing, at least one of $C_0$ and $C_I$ must be reducing (the adjusted cost of $C$ is simply the sum of adjusted costs of $C_0$ and $C_I$).
Check $\adj_{\ubar{\eps}}(C_0)$ using $\textsc{AdjCost}(\langle C_0\rangle)$.
If $C_0$ is reducing, we return the result of running the base-case algorithm on $C_0$, below.
If $C_0$ is not reducing, then $C_I$ is reducing and we return the result of recursively calling $\textsc{SimpleReducingSubcycle}(C_I)$.
If all elements of $C$ are processed without finding an intersection, then $C$ is simple and we return $C$.

There are two base cases.
First, if $C$ contains only one pathlet, then it is simple since the originating MES must also be simple.
Next, if at most two of the component pathlets of $C$ are intersecting, 
we use the following binary-search algorithm:


Given $C \coloneqq \phi_1 \circ \cdots \circ \phi_k$ where only $\phi_1$ and $\phi_k$ intersect, the \emph{only-two-intersecting} pathlets base-case is handled using a prune-and-search approach as follows.
Let $\phi_1 \coloneqq e_L \spL \Phi_1 \spoR e^+$ and $\phi_k \coloneqq e^- \spL \Phi_k \spoR e_R$ for two edges $e_L, e_R$ (initially, $e_L = e_R$), and keep an extra pointer $e_M = e_L \in \phi_1$.
We recursively shrink $\phi_1$ and $\phi_2$ until we find a simple reducing cycle, while keeping the invariant that there are no intersections between $(e_L \spL \Phi_1 \spoR e_M)$ and $(e^- \spL \Phi_k \spoR e_R)$
At each step, we query $\textsc{Intersects}((e_M \spL \Phi_1 \spoR e^+), (e^- \spL \Phi_k \spoR e_R))$.
If there is no intersection then $C$ is simple, so we return $C$.
If there is an intersection, then we invoke $\textsc{Median}(e_M \spL \Phi_1 \spoR e^+)$ to acquire $f$, the median edge of $e_M \spL \Phi_1 \spoR e^+$, and query $\textsc{Intersects}((e_M \spL \Phi_1 \spoR f), (e^- \spL \Phi_k \spoR e_R))$.
If there is no intersection up to the median, then we shrink the left pathlet to $f \spL \Phi_1 \spoR e^+$ by setting $e_M \from f$, and attempt again.
Otherwise, if $(e_M \spL \Phi_1 \spoR f) \cap (e^- \spL \Phi_k \spoR e_R) \neq \varnothing$, we invoke $\textsc{LastCommonEdge}((e_L \spL \Phi_1 \spoR f), (e^- \spL \Phi_k \spoR e_R))$ to acquire $e_1$, the \emph{last} edge of $e_L \spL \Phi_1 \spoR f$ that appears in the intersection.
There are two subcycles about $e_1$: 
\[
\begin{aligned}
\widetilde{C}_0 &\coloneqq (e_L \spL \Phi_1 \spoR e_1) \circ (e_1 \spL \Phi_k \spoR e_R) \\
\widetilde{C}_I &\coloneqq (e_1 \spL \Phi_1 \spoR e^+) \circ \phi_2 \circ \cdots \circ \phi_{k-1} \circ (e^- \spL \Phi_k \spoR e_1)
\end{aligned}
\]
In $\widetilde{C}_0$, $\abs{e_L \spL \Phi_1 \spoR e_1} < \abs{\phi_1}/2$.
In $\widetilde{C}_I$, $e_1 \spL \Phi_1 \spoR e^+$ has no intersection edges before $f$ and $\abs{f \spL \Phi_1 \spoR e^+} \le \abs{\phi_1/2}$.
Check $\adj_{\ubar{\eps}}(\widetilde{C}_0)$ using $\textsc{AdjCost}(\langle \widetilde{C}_0 \rangle)$.
If $\widetilde{C}_0$ is reducing, we restart the binary search with input $\widetilde{C}_0$.
If $\widetilde{C}_0$ is not reducing then $\widetilde{C}_I$ must be reducing, and we continue the binary search on $\widetilde{C}_I$ by setting $e_L \from e_1$, $e_R \from e_1$, and $e_M \from f$.
If $\abs{\phi_1} = 1$, then $\phi_1 = \langle e_1 \rangle$ and the intersection with $e_1$ is eliminated in both $\widetilde{C}_0$ and $\widetilde{C}_I$, so the binary search eventually terminates.

\medskip
The next lemma bounds the running time of the two procedures.
\FullVer{Let $t(n)$ be the maximum time taken by the operations of the MES data structure besides $\textsc{Report}$.
We will prove in Lemmas~\ref{L:last-edge}--\ref{L:cost} that $t(n) = O(s^6 h^4)$.
\begin{lemma}
\label{L:fte-time}
	(i) Given a compressed path $P$ in $G_\cell$, \textsc{Construct\-Expan\-sion} either returns an MES $\Phi$ of $P$ or 
	a simple reducing cycle in $O(s^4 \cdot t(n))$ time. 
	(ii) \textsc{SimpleReducingSubcycle} returns a simple reducing cycle in $O(s^4 \cdot t(n))$ time.
\end{lemma}
\begin{corollary}
	\label{C:high-level}
	The two procedures 
	$\textsc{ConstructExpansion}$ and $\textsc{Simple\-Reducing\-Sub\-cycle}$ take $O(s^{10} h^4)$ time.
\end{corollary}
}
\NotFullVer{
\begin{lemma}
\label{L:fte-time}
	(i) Given a compressed path $P$ in $G_\cell$, \textsc{Construct\-Expan\-sion} either returns an MES $\Phi$ of $P$ or 
	a simple reducing cycle in $O(s^{10} h^4)$ time. 
	(ii) \textsc{SimpleReducingSubcycle} returns a simple reducing cycle in $O(s^{10} h^4)$ time.
\end{lemma}
}

\subsection{Compact representation of pathlets}
\label{SS:pathlet_compact}

\begin{figure}
    \centering
    \includegraphics[width=0.2\textwidth, page=1]{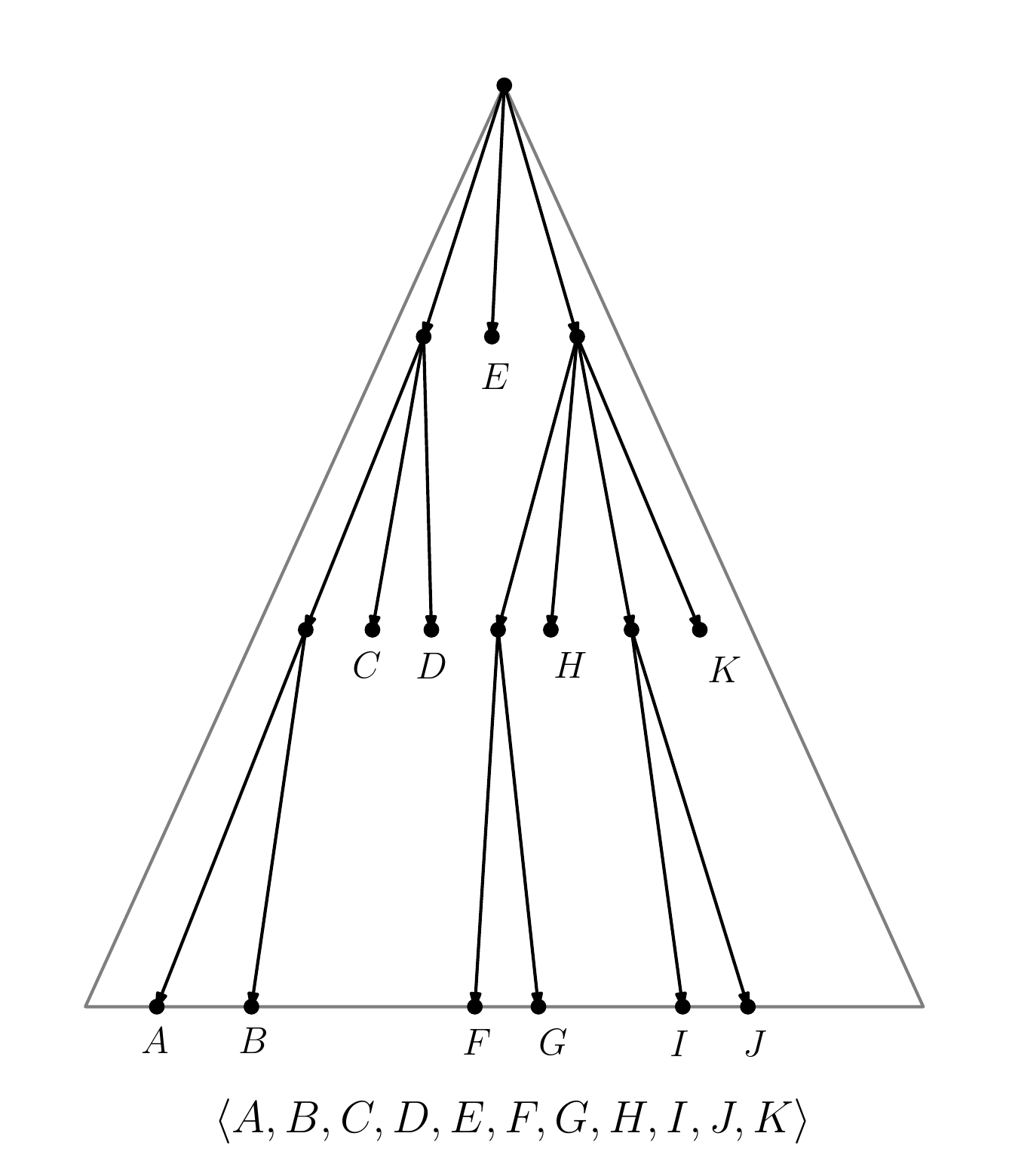}
    \includegraphics[width=0.2\textwidth, page=3]{fte}
    \caption{
        \textbf{Left}:
        An MES-tree representing a sequence $\langle A, B, C, D, E, F, G, H, I, J, K \rangle$.
        \textbf{Right}:
	Pathlet-subtree (shaded) of the same MES-tree representing subsequence 
	$\langle C, D, E, F, G, H, I \rangle$.  Its two spines are marked in red, and
        its canonical nodes are marked blue.
    }
    \label{fig:fte-tree}
\end{figure}

We now describe the data structure to store MES's and pathlets compactly.
The MES of an expansion $\Phi \coloneqq \Phi_\cell(A_\subcell, B_{\subcell'})$ is stored in a tree \EMPH{$T(\Phi)$}, called an 
\EMPH{MES tree} (to be defined next), whose leaves are the matching edges of $\Phi$, in sequence from left to right.
To reference a matching edge $e$ in $T(\Phi)$, we specify the root-leaf path to $e$ whenever $e$ is passed to a procedure or returned by a procedure.
Let the \EMPH{spine} of $T(\Phi)$ to $e$ be a root-leaf path that ends at the leaf representing $e$.
The \EMPH{pathlet subtree} of a pathlet $\phi \coloneqq e^- \spL \Phi \spR e^+$ for $e^-, e^+ \in \Phi$,
denoted by \EMPH{$T(\Phi, \phi)$}, is the subtree of $T(\Phi)$ lying between and including the two spines to $e^-$ and $e^+$.
Let $u$ be a node in $T(\Phi, \phi)$; $u$ is a node in $T(\Phi)$ as well.
If $u$ does not lie on the spines of $e^-$ or $e^+$ then all children of $u$ that appear in $T(\Phi)$ also 
appear in $T(\Phi, \phi)$; otherwise only those children of $u$ in $T(\Phi)$ that lie between the spines appear in $T(\Phi, \phi)$.
Hence, $T(\Phi, \phi)$ can be implicitly represented by storing a pointer to $T(\Phi)$ and 
spines of $e^-$ and $e^+$, which takes $O(h)=O(\log n)$ space. See Figure~\ref{fig:fte-tree}. 

MES-trees are defined recursively.
Recall that \textsc{Construct\-Expan\-sion} creates the MES for $\Phi \coloneqq \Phi_\cell(A_\subcell, B_{\subcell'})$, that has the 
form $\Phi = \phi_1 \circ \phi_2 \circ \cdots \circ \phi_t$ for some $t\le s$, where each $\phi_j$ is a pathlet of 
$\Phi(\eta_j)$ of an arc $\eta_j$ of $G_\cell$; recall that we discard empty pathlets from the concatenation.
$T(\Phi)$ consists of a root node plus $t$ subtrees $T_1, \ldots, T_t$ from left to right, where the root of $T_i$ is the $i$-th leftmost child of the root node.
If $\phi_j$ is the pathlet of $\Phi(\eta_j)$ then $T_j = T(\Phi(\eta_j), \phi_j)$.
If $\phi_j$ is a pathlet from a bridge arc then the $T_j$ is a single-node tree (matching bridge arc).
Thus, the leaves of $T(\Phi)$ are the MES-trees of (matching) bridge arcs, and all internal nodes correspond to pathlet-subtrees of internal arcs.

\mypara{Canonical pathlets and canonical nodes.}
Let $u$ be an internal non-root node of $T(\Phi)$.
The subtree of $T(\Phi)$ rooted at $u$, denoted \EMPH{$T_u$}, is a pathlet-subtree $T(\Phi(\eta), \phi)$ (which in turn is a subtree of $T(\Phi(\eta)$)), where $\eta$ is an internal arc at a descendant cell of $\cell$ and $\phi$ is a pathlet of $\Phi(\eta)$.
We call $\phi$ a \EMPH{canonical pathlet} of $\Phi(\eta)$, denoted as \EMPH{$\phi_u$}.
An MES $\Phi$ of length $k$ has $\Theta(k^2)$ possible pathlets, but we show below that the number of canonical pathlets of an expansion is much smaller (cf.\ Lemma~\ref{L:pathlet_canonical_count}).

For a pathlet $\phi$ of an expansion $\Phi$, we define the \EMPH{canonical nodes} of $T(\Phi, \phi)$ to be the ordered sequence of non-spine children of spine nodes plus the leaves at the ends of the spines, denoted by $\EMPH{$\canonical (\Phi, \phi)$} = \langle u_1, u_2, \ldots, u_k \rangle$.
$\abs{\canonical (\Phi, \phi)} = O(sh)$.
Each canonical node $u_i$ is a pathlet-subtree of some MES-tree $\Phi(\eta_i)$, i.e., represents a canonical pathlet $\phi_{u_i}$ (of $\Phi(\eta_i)$).
We can write $\phi = \phi_{u_1} \circ \cdots \circ \phi_{u_k}$ (Figure~\ref{fig:fte-tree}) --- any pathlet can be represented as the concatenation of $O(sh)$ canonical pathlets.
We associate the following auxiliary information with each node $u$ of $T(\Phi)$:
\begin{enumerate}[label=(\roman*)]\itemsep=0pt
    \item \EMPH{$b_u$} and \EMPH{$a_u$}, the first and last endpoints of $\phi_u$;
    \item $\EMPH{$m_u$} \coloneqq \abs{\phi_u}$, which is equal to the number of leaves in $T_u$;
    \item $\adj_{\ubar{\eps}}(\phi_u)$, the $\ubar{\eps}$-adjusted cost of $\phi_u$ (here we view the canonical pathlet $\phi_u$ as a path in \smash{$\vec{G}_M$} and not just a sequence of matching edges).
\end{enumerate}


\mypara{Compact form of MES-trees.}
Let $\Phi$ be an MES computed by $\textsc{ConstructExpansion}$ as 
$\Phi = \phi_1 \circ \cdots \circ \phi_t$, where pathlet $\phi_i$ originates from an MES $\Phi_i$.
Then the compact form of $T(\Phi)$ consists  of the root of $T(\Phi)$ 
plus the sequence of pathlet subtrees $T(\Phi_1,\phi_1), \ldots, T(\Phi_t,\phi_t)$, each represented by a pointer to the MES-tree and the two spines to the first and last edges.
We call the root and the spine nodes of these pathlet-subtrees the \EMPH{exposed} nodes. 
We store only the compact form of $T(\Phi)$.
For each exposed node $u$ of $T(\Phi)$, we store the 
following auxiliary information about $\phi_u$:
\begin{enumerate}[label=(\roman*)]\itemsep=0pt
    \item \EMPH{$b_u$} and \EMPH{$a_u$}, the first and last endpoints of $\phi_u$;
    \item $\EMPH{$m_u$} \coloneqq \abs{\phi_u}$, which is equal to the number of leaves in $T_u$;
    \item $\adj_{\ubar{\eps}}(\phi_u)$, the $\ubar{\eps}$-adjusted cost of $\phi_u$ (here we view the canonical pathlet $\phi_u$ as a path in \smash{$\vec{G}_M$} and not just a sequence of matching edges).
\end{enumerate}
Since there are up to $s$ children of the root, the space used by this compact representation of $T(\Phi)$  
is $O(sh)$. 
$\textsc{ConstructExpansion}$ and \textsc{SimpleReducingSubcycle} procedures return the MES of the output alternating
path/cycle in this compact form.

\begin{lemma}
	\label{L:pathlet_compact_access}
	Let $\Phi$ be an MES, and in $T(\Phi)$ let $u$ be an unexposed child of an exposed node $v$.
	Suppose the canonical pathlet $\phi_u$ (resp.\ $\phi_v$) originates from  $\Phi_u$ (resp.\ $\Phi_v$).
	Let $u'$ be the child of the root of $T(\Phi_v, \phi_v)$ corresponding to the same arc as $u$.
	Then $T(\Phi_u,\phi_u)$ is identical to the subtree at $u' \in T(\Phi_v)$.
	%
\end{lemma}

Lemma~\ref{L:pathlet_compact_access} suggests how we can traverse $T(\Phi)$ using its compact form.
An exposed node of $T(\Phi)$ can be reached by following the provided spine.
Suppose we want to traverse into an unexposed child $u$ of an exposed node $v$ in $T(\Phi)$.
Then we query the compact form of $T(\Phi_v)$, where (a copy of) $u$ is an exposed node, and recurse from $T(\Phi_v)$.
Since $u$ is unexposed, none of the spines from $T(\Phi)$ include $u$ or its subtree.

\mypara{Cells, level, and rank.}
Let $\Phi$ be an expansion in a cell $\cell$ of level $i$.
We assign the \emph{cell} and \emph{level} of $\Phi$ to be $\cell$ and $i$, respectively.
Recall that expansion $\Phi(\gamma)$ of an internal arc $\gamma$ at $\cell$ is an expansion from a child cell of $\cell$ (given by its weight certificate) --- $\Phi(\gamma)$ inherits cell and level assigned to that expansion.
If $\gamma$ is a bridge arc at a cell $\Delta$, then we assign $\Delta$ and the level of $\Delta$ as the cell and level of $\Phi(\gamma)$, respectively.
For a pathlet $\phi$ originating from $\Phi$, we assign the cell and level of $\phi$ to be those of $\Phi$.
For a node $u$ of $T(\Phi)$, we define the cell of $u$, denoted by \EMPH{$\cell(u)$}, to be the cell 
assigned to the canonical pathlet $\phi_u$ associated with $u$, and \EMPH{$\level(u)$} to be the level of $\cell(u)$.
Hence, if a child $u$ of an internal node $v$ in $T(\Phi)$ is an internal node (resp.\ leaf) then 
$\level(u) = \level(v)-1$ (resp. $\level(u)=\level(v)$).

The \EMPH{rank} of a canonical pathlet $\phi$ originating from $\Phi$, denoted \EMPH{$\rank(\phi)$}, is the smallest 
value $i$ such that there is a node $u$ in an MES-tree $T(\Phi')$ for some level-$i$ expansion $\Phi'$ with $T(\Phi,\phi)$ being identified with 
the subtree of $T(\Phi')$ rooted at $u$, i.e., $\phi=\phi_u$.

\begin{lemma}
\label{L:pathlet_canonical_exposed}
Let $\phi_u$ be a canonical pathlet associated with a node $u$ of an MES-tree $T(\Phi)$.
	Then there is another MES $\Phi'$ with $\level(\Phi') \le \level(\Phi)$ and a node $u'$ in $T(\Phi')$ such that $\phi_u=\phi_{u'}$ and $u'$ is an exposed node of $T(\Phi')$.
\end{lemma}

The above lemma helps to prove Lemma~\ref{L:pathlet_canonical_count}, which bounds the number of canonical pathlets.
\begin{lemma}
	\label{L:pathlet_canonical_count}
		(i) There are $O(n s^3 h^2)$ canonical pathlets.
		(ii) For any cell $\cell \in \Cover$, there are $O(s^3 h)$ canonical pathlets $\phi$ whose cell is $\cell$.
\end{lemma}

\FullVer{
\begin{proof}
First we prove (i).
By Lemma~\ref{L:pathlet_canonical_exposed}, any canonical pathlet can be identified with an exposed node of an MES-tree.
An MES-tree has $O(sh)$ exposed nodes, there are $O(s^2)$ MES's comping from expansions at a cell, and each level has $O(n)$ nonempty cells, so the bound follows.

Next we prove (ii).
Let $\phi$ be a canonical pathlet with $\cell(\phi) = \cell$.
Then there is an MES $\Phi$ at an ancestor cell of $\cell$ such that $\phi$ is associated with an exposed node of $T(\Phi)$.
Each spine of $T(\Phi)$ has at most one node $u$ where $\cell(u) = \cell$; $T(\Phi)$ has $O(s)$ spines; $\cell$ has $O(h)$ ancestor cells, each having $O(s^2)$ MES's; the bound follows.
\end{proof}
}

\mypara{Intersection tables.}
Define $\GLOS[canonical-pathlets]{Set of canonical pathlets originate from $\Phi$}{\mathcal{P}(\Phi)}$ as the set of canonical pathlets that originate from $\Phi$.
Let $\eta_1, \eta_2$ be two arcs of a pair of sibling cells.
Recall that if $\eta$ is a matching bridge arc then $\Phi(\eta)$ is the longest matching edge corresponding to $\eta$.
We maintain an \EMPH{intersection table} $\GLOS[intersection-table]{Intersection table for all $(\phi_1, \phi_2) \in \mathcal{P}(\Phi(\eta_1)) \times \mathcal{P}(\Phi(\eta_2))$}{\beta_{\eta_1, \eta_2}}: \mathcal{P}(\Phi(\eta_1)) \times \mathcal{P}(\Phi(\eta_2)) \to \{0,1\}$ such that
$\beta_{\eta_1, \eta_2}(\phi_1, \phi_2) = 1$ if the two pathlets share an edge and $0$ otherwise.

The intersection table is populated dynamically.
In particular, we maintain the following invariant:
before we compute MES for any cells at level $i$, we have already computed $\beta_{\eta_1, \eta_2}(\phi_1, \phi_2)$ for all pairs of pathlets of rank at most $i-1$.
Therefore, after we finish computing MES for cells at level $i$, we compute all not-yet-computed $\beta_{\eta_1, \eta_2}(\phi_1, \phi_2)$ where $\phi_1, \phi_2$ are canonical pathlets of rank at most $i$ (i.e., one of $\phi_1, \phi_2$ has rank $i$), using the intersection query $\textsc{Intersection}$ we present later in this section.

Since $\abs{\mathcal{P}(\Phi(\eta))} = O(s^3 h)$ and the total number of canonical pathlets is $O(n s^3 h^2)$ by Lemma~\ref{L:pathlet_canonical_count}, the total size of intersection tables is $O(n s^6 h^3)$.
We thus obtain the following.

\begin{lemma}
The total size of the data structure that maintains MES is $O(n s^6 h^3)$, where $h$ is the height of the hierarchy and $s$ is the maximum number of clusters in a cell $\cell$.
\end{lemma}

\subsection{Pathlet operations}
\label{SS:pathlet_ops}

We now describe how operations on pathlets and expansions (cf.\ Section~\ref{SS:pathlet})
are implemented using MES and pathlet trees.


\mypara{$\textsc{Intersects}$.}
Given two pathlets $\phi_1, \phi_2$ originating from
lower-level MES's $\Phi_1=\Phi(\eta_1), \Phi_2=\Phi(\eta_2)$ (which are expansions of descendant cells of $\cell$)
respectively, the \textsc{Intersects} procedure determines whether  
$\phi_1 \cap \phi_2 \neq \varnothing$.
The input pathlets $\phi_1$ and $\phi_2$ are given as (the spines of) pathlet-subtrees $T(\Phi_1, \phi_1)$ and $T(\Phi_2, \phi_2)$.
For $i=1,2$, let $\canonical(\Phi_i,\phi_i)$ be the set of canonical nodes in of $T(\Phi_i,\phi_i)$.
We describe the intersection-detection procedure for a pair $(u_1,u_2) \in \canonical (\Phi_1,\phi_1)\times \canonical (\Phi_2,\phi_2)$.
Using this procedure for all such pairs, we determine whether $\phi_1 \cap \phi_2 \neq \varnothing$.

Fix a pair $u_1 \in \canonical (\Phi_1, \phi_1)$ and $u_2 \in \canonical (\Phi_2, \phi_2)$. We want to determine whether $\phi_{u_1}\cap\phi_{u_2} \ne\varnothing$.
If $\cell(u_1) \cap \cell(u_2) =\varnothing$ then $\phi_{u_1} \cap \phi_{u_2}=\varnothing$, so we return no.
Assume that $\cell(u_1) \cap \cell(u_2) \neq \varnothing$.
If $\level(u_1)=\level(u_2)$, then $\cell(u_1), \cell(u_2)$ are sibling cells and $\eta_i$ is an arc in $G_{\cell(u_i)}$.
If $\eta_1$ or $\eta_2$ is a bridge arc then we return yes only if $\eta_1$ and $\eta_2$ are the same 
matching edge, otherwise return no. On the other hand, if $\eta_1, \eta_2$ are internal arcs then both
$\level (\eta_i)$ and  $\rank(\phi_i)$ is less than the level of $\cell$,
so $\beta_{\eta_1,\eta_2} (\phi_{u_1},\phi_{u_2})$ has 
been computed and we return this value.
%

We now assume 
that $\level(u_1) > \level(u_2)$ without loss of generality.
If $u_1$ is a leaf, then it is impossible for the matching bridge edge $\phi_{u_1}$
to appear in $u_2$ (which is composed of lower-level edges) so we return no.
If $u_1$ is an internal node then we recursively test $u_2$ against every children $v$ of $u_1$, to determine whether 
$\phi_{v}\cap\phi_{u_2} \ne\varnothing$. If any of them returns yes, we return yes. 
The following lemma bounds the running time using a packing argument.

\begin{lemma}
	\label{L:canonical-intersect}
	Given a pair of pathlets $\phi_1, \phi_2$ originating from $\Phi_1,\Phi_2$,
	for a fixed pair of canonical nodes $(u_1, u_2) \in\canonical (\Phi_1,\phi_1)\times\canonical (\Phi_2,\phi_2)$,
	\textsc{Intersects} takes $O(s^4h^2)$ time.
\end{lemma}

\FullVer{
\begin{proof}
	If $\level(u_1) =\level(u_2)$ or $\cell(u_1)\cap\cell(u_2)=\varnothing$, then the lemma obviously holds because the
	procedure performs at most one table look up and no recursive call is made.
	Without loss of generality, assume that $\level(u_1)>\level(u_2)$, and let $T_{u_1}$ be the subtree of $T(\Phi_1)$ rooted at $u_1$.
	Note that for any child $v$ of $u_1$, $\level(v)\ge \level(u_1)-1 \ge \level(u_2)$, the algorithm never 
	recursively explores the children of $u_2$ and no further recursive calls are made when the procedure 
	reaches a descendant of $u_1$ of $\level(u_2)$.
	For any fixed level $i\ge\level(u_2)$, there are $O(1)$ cells
	$\Delta\in\C$ such that $\Delta\cap\cell (u_2) \ne \varnothing$. By Lemma~\ref{L:pathlet_canonical_count}~(ii), there are $O(s^3 h)$
	canonical pathlets $\varphi$ with $\cell(\varphi)=\Delta$. Hence, the procedure recursively visits the children of 
	 $O(s^3h)$ nodes of $T_{u_1}$ of level $i$.
	 Since the height of $T_{u_1}$ is at most $h$ and the maximum degree of a node in $T_{u_1}$ is $s$,
	the lemma follows.
\end{proof}
}
Summing this bound over all pairs of canonical nodes, we obtain the following:
\begin{corollary}
	\label{C:intersect}
	\textsc{Intersects} procedure takes $O(s^6h^4)$ time.
\end{corollary}


\mypara{$\textsc{LastCommonEdge}$.}
Given two pathlets $\phi_1, \phi_2$ originating from lower-level MES's $\Phi_1=\Phi(\eta_1), \Phi_2=\Phi(\eta_2)$,
respectively, returns the \emph{last} edge of $\phi_1$ that appears in $\phi_2$.  If such an edge exists, let $e_1$ denote this edge.
The procedure first finds the edge $e_1\in\phi_1$, then its copy in $\phi_2$, and then computes the spines of $e_1$, of
its copy in $\phi_2$, and of their predecessors.

By invoking \textsc{Intersects}, we first determine whether $\phi_1\cap\phi_2\ne\varnothing$. If the answer is yes, 
we also find the last canonical node $u_1\in\canonical(\Phi_1,\phi_1)$ whose canonical pathlet $\phi_{u_1}$
that contains $e_1$. 
Let $U_2 \coloneqq \{u_2\in\canonical(\Phi_2,\phi_2) \mid \phi_{u_1}\cap\phi_{u_2}\ne\varnothing\}$.
%
If $u_1$ is a leaf node with matching edge $e$, then we set $e_1 \from e$. We assume that we have the spine 
to $u_1$ in $\Phi_1$.
If $u_1$ is not a leaf, then by invoking \textsc{Intersects} for all pairs $\childr(u_1)\times U_2$ (where $\childr(u_1)$ is the set of children of $u_1$ in $T(\Phi_1)$), 
we find the last child $v\in\childr(u_1)$ that intersects the canonical pathlet corresponding to a node in $U_2$. 
We now set $u_1 \from v$ and $U_2$ as above.
Let $U_1$ be the children of $u_1$ in $T(\Phi_1,\phi_1)$ and $U_2$ to be the subset of nodes $u_2\in U_2$ such that $\phi_{u_1}\cap\phi_{u_2} \ne\varnothing$. 

After having computed $e_1$, we compute its copy in $\phi_2$ as follows. Let $u_2\in U_2$ be the node such that $e_1\in\phi_{u_2}$. Since $\phi_2$ does not have any duplicate edges, $u_2$ is unique.
By following a variant of \textsc{Intersects} procedure, we can traverse to the unique leaf of $T(\Phi_2,\phi_2)$ that contains a copy of $e_1$.
After computing spines for $e_1$ and $e_2$, the predecessors $e_3, e_4$ can be computed by traversing backwards up the spines to find the first leaf left of $e_1, e_2$ respectively.
If $e_1$ (resp.\ $e_2$) happen to be the first edge in $\phi_1$ (resp.\ $\phi_2$), then we return $e_3$ (resp.\ $e_4$) as \textit{null} instead.

Since the maximum degree of a node in $T(\Phi_1)$ is $s$, the procedure calls \textsc{Intersects} procedure for $O(s^2 h)$ pairs of canonical nodes at each level of the recursion. The depth of the recursion is $h$, so by Lemma~\ref{L:canonical-intersect}, we obtain the following:
\begin{lemma}
	\label{L:last-edge}
	$\textsc{LastCommonEdge}$ runs in time $O(s^6h^4)$.
\end{lemma}

\mypara{$\textsc{Median}$.}
Let $\phi$ be a pathlet originating from an MES $\Phi$, represented as the spine of $T(\Phi,\phi)$.
Using the length information stored at the nodes of $T(\Phi)$, the spine of the median edge of $\phi$ can be computed in $O(sh)$ time by traversing $T(\Phi, \phi)$ in a top-down manner.

\begin{lemma}
	\label{L:median}
	$\textsc{Median}$ runs in time $O(sh)$.
\end{lemma}

\mypara{$\textsc{AdjCost}$.}
Given the compact representation $\langle \Pi\rangle$ of a path/cycle $\Pi$ in \smash{$\vec{G}_M$}, whose
MES is represented as a sequence $\phi_1\circ\cdots\circ\phi_t$ of at most $s$ pathlets,
the procedure returns $\adj_{\ubar{\eps}} (\Pi)$. Let $b_i$ and $a_i$ be the starting and ending tips of $\phi_i$ 
respectively, and suppose $\phi_i$ originates from the MES $\Phi_i$. Since the compact representation of $\phi_i$ consists of the spines of $T(\Phi_i,\phi_i)$, by accessing the nodes of $\canonical (\Phi_i,\phi_i)$, we can compute 
$\adj_{\ubar{\eps}}(\phi_i)$ in $O(sh)$ time. Next we compute 
$$\theta \coloneqq \adj_{\ubar{\eps}} (\phi_1\circ\cdots\circ\phi_t) = \sum_{i=1}^t \adj_{\ubar{\eps}}(\phi_i) + \sum_{i=1}^{t-1} \adj_{\ubar{\eps}}(a_i b_{i+1}).$$

If $\Pi$ is a cycle then we set $\adj_{\ubar{\eps}} (\Pi) \coloneqq \theta+\adj_{\ubar{\eps}}(a_tb_1)$. 
If $\Pi$ is a path with its beginning and ending tips being $b_1$ and $a_t$ respectively, then we set $\adj_{\ubar{\eps}} (\Pi) \coloneqq \theta$. Finally, if the first and last edges of $\Pi$ are non-matching edges with its starting and ending tips being $a_0$ and $b_{t+1}$, respectively, then we set $\adj_{\ubar{\eps}} (\Pi) \coloneqq \theta +\adj_{\ubar{\eps}} (a_0b_1)+\adj_{\ubar{\eps}} (a_tb_{t+1})$. 

\begin{lemma}
	\label{L:cost}
	Given the compact representation of a path or a cycle in $\vec{G}_M$ with its MES composed of at most $s$ pathlets,
	$\textsc{AdjCost}$ runs in time $O(s^2h)$.
\end{lemma}

\mypara{$\textsc{Report}$.}
Given the compact representation of a path/cycle $\langle \Pi\rangle$ of a path/cycle $\Pi$ in \smash{$\vec{G}_M$}, whose
MES is represented as a sequence $\phi_1\circ\cdots\circ\phi_t$ of at most $s$ pathlets, the procedure returns the sequence of edges in $\Pi$.
Let $b_i, a_i, \Phi_i$ be as above. 

We first compute the sequence of edges in the alternating path $\phi_1\circ\cdots\circ\phi_t$ with $b_1$ and $a_t$ as its tips.  
If $t=1$ and $T(\Phi_1,\phi_1)$ is a leaf, then the desired path is the matching (bridge) edge associated with the leaf.
If $t=1$ but $\phi_1$ consists of more than one edge,
then we recursively traverse the subtree rooted at each canonical node of $\canonical (\Phi_1,\phi_1)$, compute the 
alternating path represented by each canonical pathlet of $\phi_1$,
concatenate the results in sequence, and add the non-matching edges between two canonical pathlets.
For $t>1$, we repeat this procedure for each $\phi_i$,  concatenate the paths for all $\phi_i$'s,
and add the non-matching edges between the paths of two consecutive pathlets. Let $\Gamma$ be the resulting 
sequence of edges. 

If $\Pi$ is an alternating path from $b_1$ to $a_t$, we simply return $\Gamma$. If $\Pi$ is a path with its starting and ending tips being $a_0$ and $b_{t+1}$, we return $a_0b_1 \circ \Gamma \circ a_tb_{t+1}$. Finally, if $\Pi$ is a cycle then we return the cycle $\Gamma\circ b_ta_1$.

\begin{lemma}
	\label{L:report}
	Given the compact representation of a path or a cycle in \smash{$\vec{G}_M$} with its MES composed of at most $s$ 
	pathlets,
	$\textsc{Report}$ run in time $O((s^2+k)h)$ where $k$ is the output size.
\end{lemma}

\mypara{$\textsc{Splice}$ and $\textsc{Concatenate}$.}
Assuming that an edge $e$ of a pathlet $\phi$, originating from an MES $\Phi$, is specified by the root-to-leaf path in $T(\Phi,\phi)$. 
The splice operation $e\spL\phi$ can be performed in $O(sh)$ time by creating a pathlet subtree for the spliced pathlet from the compact representation of $T(\Phi,\phi)$.

Given pathlet subtrees $T(\Phi_1,\phi_1), \ldots, T(\Phi_t,\phi_t)$ of $\phi_1, \ldots, \phi_t$ 
in their compact forms, the MES tree $T(\Phi)$ for $\Phi = \phi_1 \circ \cdots \circ \phi_t$ can be constructed in $O(sh)$ time by first creating the root of $T(\Phi)$, then
copying the spines of each $\phi_j$, and attaching it as the subtree rooted at the $j$-th child of the root.
We omit the straightforward details of the two procedures.

\mypara{Updating intersection tables.}
After we have computed MES's of all affected cells at level $i$, we update the intersection tables to add the entries for canonical
pathlets of rank $i$. Let $\Phi$ be an MES in a cell $\cell$ at level $i$ that was newly constructed. The canonical pathlets associated with the exposed nodes of $T(\Phi)$ are new canonical pathlets for which the intersection-table entries need to be computed.
Let $\varphi_1$ be such a canonical pathlet originating from an MES $\Phi_{\eta_1}$. Let $\Delta_1 \coloneqq \cell(\Phi_{\eta_1})$.
Then for all matching/internal arcs $\eta_2$ in sibling cells $\Delta_2$ of $\Delta_1$ and for all canonical pathlets of $\Phi(\eta_2)$ of rank at most $i$ (which we already have at our disposal), we compute $\beta_{\eta_1,\eta_2} (\varphi_1,\varphi_2)$ by calling \textsc{Intersects} ($\varphi_1$, $\varphi_2$).
Since there are $O(1)$ sibling cells of $\Delta_1$, each with $O(s^2)$ arcs, and there are $O(s^3 h)$ canonical pathlets originating from each arc, the total number of entries computed for $\varphi_1$ is $O(s^5 h)$.
MES $\Phi$ has $O(sh)$ canonical pathlets, and $\cell$ has $O(s^2)$ MES's, so the total number entries computed for $\cell$ is $O(s^8h^2)$. 
Using Corollary~\ref{C:intersect}, we obtain the following:
\begin{lemma}
	\label{L:table-lookup}
	The total time spent in updating the intersection tables because of the new canonical pathlets generated in the computation of expansions at a cell is $O(s^{14}h^6)$.
\end{lemma}

Putting everything together, we obtain the following:

\begin{lemma}
	\label{L:MESs}
	For any cell $\cell$ at level $i$, the total time spent in computing expansions, constructing MES-trees, and updating intersection tables, assuming that 
	all nodes at level greater than $i$  have been been processed,
	is $(\eps^{-1}\log n)^{O(d)}$. 
\end{lemma}

\section{\textsc{FindPath}, \textsc{Augment}, and \textsc{Repair} Procedures}
\label{S:ds_procedures}

We describe the operations $\textsc{FindPath}$, $\textsc{Augment}$, and $\textsc{Repair}$ performed on the data structure built on each cell $\cell$.
For an alternating path $\Pi$ in $\vec{G}_M$, we call a cell $\cell \in \Cover$ \EMPH{affected} by~$\Pi$ if $\cell$ contains a vertex of $\Pi$ (that is, a point in $\reals^d$).
Let $\GLOS[affected-cells]{Cells affected by $\Pi$}{\Cover_\Pi} \subseteq \Cover$ denote the set of cells affected by $\Pi$.

\mypara{$\textsc{FindPath}()$:}
It performs a $\textsc{DeleteMin}$ operation on the priority queue $\mathcal{Q}$ storing the candidate subcell pairs $\Pairs$ and 
retrieves a pair $(\subcell_\cell, \subcell_\cell')$, such that 
$\Phi \coloneqq \Phi_\cell(A_{\subcell_\cell}, B_{\subcell_\cell'})$ is an augmenting path.
Recall that the data structure at $\cell$ stores a compact representation $\langle\Phi\rangle$ of $\Phi$.
The procedure calls $\textsc{Report} (\langle\Phi\rangle)$ and returns the resulting sequences of edges.
The total time spent is $\abs{\Phi} \cdot (\eps^{-1}\log n)^{O(d)}$.

\mypara{$\textsc{Augment}(M, \Pi)$:}
It first sets $M \from M \oplus \Pi$.
We store $\Cover_\Pi$, the set of cells affected by $\Pi$, in a priority queue $\Xi$ with the level of the cell as its key.
At each step, we retrieve a cell $\cell$ from $\Xi$ of the lowest level and update the data structure by calling $\textsc{Repair}(\cell)$.
If the call to $\textsc{Repair}$ returns a reducing simple cycle $\Gamma$, we set $M \from M \oplus \Gamma$ and insert all cells in $\Cover_\Gamma$ into $\Xi$ and continue.
This process continues until $\Xi$ becomes empty.
The procedure then returns $M$.

Let $\Gamma_1, \ldots, \Gamma_t$ be the reducing cycles returned by $\textsc{Repair}$.
As we see below, $\textsc{Repair}$ takes $(\eps^{-1} \log n)^{O(d)}$ amortized time.
Then $\textsc{Augment}$ takes time $\Paren{\abs{\Pi} + \sum_i \abs{\Gamma_i}} \cdot (\eps^{-1} \log n)^{O(d)}$ time.

\begin{figure}[h!]
\small
\centering\begin{algorithm}
\textul{$\textsc{Augment}(M, \Pi)$:}\+
\\  $\Xi \from \varnothing$,\quad $M \from M \oplus \Pi$
\\  for all $\cell \in \Cover_\Pi$: \;
        $\textsc{Insert}(\cell, \Xi)$ 
\\  while $\Xi \neq \varnothing$:\+
\\  $\cell \from \textsc{DeleteMin}(\Xi)$,\quad
    $\Gamma \from \textsc{Repair}(\cell)$
\\  if $\Gamma \neq \textit{null}$:\+
\\      $M \from M \oplus \Gamma$
\\      for all $\cell' \in \Cover_\Gamma$: \;
            $\textsc{Insert}(\cell', \Xi)$
\-\-
\\ return $M$
\end{algorithm}
\caption{Implementation of \textsc{Augment} using \textsc{Repair}.}
\label{F:augment}
\end{figure}

\mypara{$\textsc{Repair}(\cell)$:}
$\textsc{Repair}$ is always called in a bottom-up manner, so we assume that all children of $\cell$ that were affected by the update in $M$ are stable before the invocation of $\textsc{Repair}$ at $\cell$.
$\textsc{Repair}(\cell)$ reconstructs all the information stored at $\cell$, in the following steps:
\begin{enumerate}[label=(\roman*), ref=(\roman*)]
    \item \label{it:repair_1}
		(\textbf{Updating clusters})
        For each subcell $\subcell$ of $\cell$, we update the saturation status of $A_\subcell$ and $B_\subcell$.
        We maintain the free vertices of $A_\subcell$ and $B_\subcell$ in
        linked lists, and as they get matched we delete them.

    \item \label{it:repair_2}
		(\textbf{Assigning arc weights})
        Let $\subcell, \subcell'$ be subcells of $\cell$.
        \begin{itemize}\itemsep=0pt
        \item
        If no child cell of $\cell$ contains both $\subcell$ and $\subcell'$, then we update the set of matching edges 
		whose endpoints are in $\subcell, \subcell'$ --- delete the edges that are no longer in $M$ and insert newly created edges of $M$ whose endpoints lie in $\subcell$ and $\subcell'$.
		We use Eq.~(\ref{E:weight_nonmatch_bridge})
		and (\ref{E:weight_match_bridge})
		to compute the weights of arcs between the clusters of
		$\subcell$ and $\subcell'$.

        \item
        If some child cell contains both $\subcell$ and $\subcell'$, we use the recursive expression
		in Eq.~(\ref{E:weight_match_internal}) 
		to compute the weights of
		arcs between the clusters of $\subcell, \subcell'$.
        \end{itemize}

    \item \label{it:repair_3}
		(\textbf{APSP computation})
        We compute the minimum-weight compressed paths between every pair of nodes in $G_\cell$.
		If during this computation, we find a negative cycle $C$, (i.e.\ $w_\cell(C) < 0$), we compute (the compact 
		representation of) a simple reducing subcycle $\Gamma$
		by calling \textsc{SimpleReducingSubCycle}($C$).
        In this case, we abort the update of $\cell$ and return 
        $\textsc{Report}(\langle\Gamma\rangle)$,
        which retrieves the sequence of edges in $\Gamma$.

    \item \label{it:repair_4}
		(\textbf{Computing expansions})
        For every pair of subcells $\subcell, \subcell'$, we compute $\Phi_\cell(A_\subcell, B_{\subcell'})$ as described above.
        The procedure may abort and return a simple reducing cycle $\Gamma$, in which case we abort the update of 
		$\cell$ and return $\textsc{Report}(\langle\Gamma\rangle)$.
        If we succeed in computing expansions for all pairs, $\cell$ is stable.

    \item \label{it:repair_5}
		(\textbf{Augmenting paths})
        For each pair of subcells $\subcell, \subcell'$ of $\cell$ such that both $A_\subcell$ and $B_{\subcell'}$ are 
		unsaturated, let $\Pi_{\subcell,\subcell'} := \Phi_\cell(\subcell, \subcell')$; by construction, 
		its tips are free points. We computing $\adj_{\ubar{\eps}}(\Pi_{\subcell, \subcell'})$ by calling
		$\textsc{AdjCost} (\langle \Pi_{\subcell,\subcell'}\rangle)$.
        Otherwise, $\Pi_{\subcell, \subcell'}$ is undefined and $\adj_{\ubar{\eps}}(\Pi_{\subcell, \subcell'}) = \infty$.
        We compute $(\subcell_\cell, \subcell_\cell') \coloneqq \argmin_{\subcell, \subcell'} \adj_{\ubar{\eps}}(\Pi_{\subcell, \subcell'})$ and
        insert $(\subcell_\cell, \subcell_\cell')$ into $\Pairs$.
\end{enumerate}

\FullVer{
The time spent in Steps~\ref{it:repair_1} and~\ref{it:repair_2} is charged to changes in the current matching.
Each matching edge that is deleted or inserted is charged $O(\log n)$ times --- $O(1)$ times per level.
If Step~\ref{it:repair_3} or~\ref{it:repair_4} calls $\textsc{Report}(\langle\Gamma\rangle)$, it takes $\abs{\Gamma} \cdot (\eps^{-1}\log n)^{O(d)}$ time, which we charge to the cycle returned.
Hence, the total charge to the matching edges and reducing cycles over all calls of $\textsc{Repair}$ during a single execution of $\textsc{Augment}$ is $\Paren{\sum_i \abs{\Gamma_i} + \abs{\Pi}} \cdot (\eps^{-1}\log n)^{O(d)}$, which we charge to $\textsc{Augment}$.
All other operations, including computing the compressed paths in $G_\cell$ and constructing expansion-trees take $(\eps^{-1} \log n)^{O(d)}$ time.
Hence, we conclude:
}
\NotFullVer{
	Omitting the details, we conclude:
}

\begin{lemma}
\label{L:procedure_times}
$\textsc{FindPath}$ takes $\abs{\Pi} \cdot (\eps^{-1}\log n)^{O(d)}$ time, where $\Pi$ is the augmenting path returned by the procedure.
$\textsc{Augment} (\Pi, M)$ takes $\Paren{\abs{\Pi} + \sum_i \abs{\Gamma_i}} \cdot (\eps^{-1} \log n)^{O(d)}$ time, where $\Set{\Gamma_i}$ is the set of reducing cycles canceled by the procedure.
\end{lemma}

The correctness of $\textsc{FindPath}$ and $\textsc{Augment}$ and the invariant~\ref{I:cycle-invariant} follow from the following lemmas:
\begin{lemma}
\label{L:findpath}
$\textsc{FindPath}$ returns an augmenting path $\Pi$ such that $\adj_{\ubar{\eps}}(\Pi) \le \adj^*_{\bar{\eps},M}$.
\end{lemma}
\begin{proof}
Let $M$ be the current matching in the beginning of an iteration (when $\textsc{FindPath}$ is called), and let $\Pi^*$ be the augmenting path in $\vec{G}_M$ with the minimum $\bar{\eps}$-adjusted cost, i.e., $\adj_{\bar{\eps}}(\Pi^*) = \adj^*_{\bar{\eps},M}$.
By Lemma~\ref{L:container_exists}, there is a cell in $\Cover$ that contains~$\Pi^*$.
Let $\cell$ be the smallest cell in $\Cover$ that contains $\Pi^*$ (if there is more than one, choose an arbitrary one).
	Let $\subcell$ (resp.\ $\subcell'$) be the subcell of $\cell$ that contains the starting (resp.\ ending) tip of $\Pi^*$.
By Lifting Inequality (Lemma~\ref{L:lift}),
\[
    w_\cell(\pi_\cell(A_\subcell, B_{\subcell'}))
	\le \adj_{\bar{\eps}}(\Pi^*) = \adj_{\bar{\eps}}^*.
\]
	Since both $A_\subcell$ and $B_{\subcell'}$ are unsaturated, $\textsc{Repair}$ procedure considers the 
	expansion $\Phi_\cell(A_\subcell,B_{\subcell'})$ with free vertices being its tips as a candidate augmenting path.
By Expansion Inequality (Lemma~\ref{L:expansion}),
\begin{equation}
    \adj_{\ubar{\eps}}(\Phi_\cell(A_\subcell,B_{\subcell'}))
        \le w_\cell(\pi_\cell(A_\subcell, B_{\subcell'})).
\end{equation}
Let $(\subcell_\cell, \subcell_\cell')$ be the pair chosen by $\textsc{Repair}(\cell)$ for $\cell$ 
	in Step~\ref{it:repair_5}: 
\[
	\adj_{\ubar{\eps}}(\Phi_\cell(A_{\subcell_\cell},B_{\subcell'_\cell}))
	\le \adj_{\ubar{\eps}}(\Phi_\cell(A_\subcell,B_{\subcell'}))
        \le \adj^*_{\bar\e}.
\]
Since $(\subcell_\cell, \subcell_\cell') \in \Pairs$, 
	$\textsc{FindPath}$ returns an augmenting path $\Pi$ with 
	$\adj_{\ubar{\eps}}(\Pi) \le \adj_{\ubar{\eps}}(\Phi_\cell(A_{\subcell_\cell},B_{\subcell'_\cell})) \le \adj^*_{\bar\e}$.
\end{proof}

\begin{lemma}
\label{L:augment_reducingcycle}
Let $M$ be any matching.
If $\vec{G}_M$ contains an alternating cycle $\Gamma$ with $\adj_{\bar{\eps}}(\Gamma) < 0$ then $\textsc{Augment}$ returns a reducing cycle.
\end{lemma}

\FullVer{
\begin{proof}
Suppose $\vec{G}_M$ contains a cycle $\Gamma$ with $\adj_{\bar{\eps}}(\Gamma) < 0$.
Then, as in the proof of Lemma~\ref{L:findpath}, we can apply Lemma~\ref{L:lift} to argue that either the compressed graph $G_\cell$ at the smallest cell $\cell \in \Cover$ containing $\Gamma$ contains a negative cycle $C$, or one of the descendants of $\cell$ is not stable and therefore a descendant of $\cell$ contains a negative compressed cycle $C$.
By Corollary~\ref{C:negativecycle_reducingcycle}, the expansion of $C$ is a reducing cycle, and 
	\textsc{Repair} returns a simple reducing subcycle of the expansion of $C$.
\end{proof}
}

\begin{corollary}
\label{C:cycleinv}
Invariant~\ref{I:cycle-invariant} holds at the start of each iteration.
\end{corollary}


\section{Analysis of the Algorithm}
\label{S:analysis}

In this section, we analyze the performance of the overall algorithm.
We first analyze the cost of $M_\alg$, the matching computed by the algorithm, and then analyze the running time of the algorithm.

\begin{lemma}
\label{L:intermediate_augpath_full}
The following properties hold throughout the algorithm:
    (i) The cost of any intermediate matching is at most $(1+\frac{\eps}{2}) \cdot \cost(M_\opt)$.
    (ii) In the beginning of each iteration when the invariant~\ref{I:cycle-invariant} is satisfied, let $M$ be the 
	    current matching and $\Pi$ any augmenting path (with respect to $M$) such that 
		$\adj_{\ubar{\eps}}(\Pi) \le \adj^*_{\bar{\eps},M}$.
    Then,
    \[
        \net\cost(\Pi) \le (1+\frac{\eps}{2}) \cdot \cost(M_\opt) -  \cost(M).
    \]
\end{lemma}
\begin{proof}
We prove the two statements of the lemma together by induction on the number of iterations.
Initially $M = \varnothing$ and each edge of $M_\opt$ is an augmenting path, so both claims hold trivially.
Suppose the claims hold for the first $i-1$ iterations.
Consider the $i$-th iteration, and let $M$ be the current matching at the beginning of the $i$-th iteration.
By induction hypothesis, $\cost(M) \le (1+\frac{\eps}{2}) \cdot \cost(M_\opt)$.

If $\adj^*_{\bar{\eps}} < 0$, then
\[
\net\cost(\Pi)
    \le \adj_{\ubar{\eps}}(\Pi)
    \le \adj^*_{\bar{\eps}}
    < 0
    \le (1+\frac{\eps}{2})\cdot \cost(M_\opt) - \cost(M).
\]
Since $\textsc{FindPath}$ returns an augmenting path $P$ with $\net\cost(P) \le \adj_{\ubar{\eps}}(P) \le \adj^*_{\bar{\eps}} < 0$,
	\begin{equation}
		\label{eq:path-cost1}
\cost(M \oplus P)
    = \cost(M) + \net\cost(P)
    \le \cost(M)
    \le (1+\frac{\eps}{2}) \cdot \cost(M_\opt).
	\end{equation}

Next, consider the case when $\adj^* \ge 0$.
The symmetric difference $M \oplus M_\opt$ consists of a nonempty set $\mathcal{P}$ of pairwise-disjoint augmenting paths and a (possibly empty) set $\mathcal{N}$ of alternating cycles.
Every augmenting path $\Pi' \in \mathcal{P}$ has $\adj_{\bar{\eps}}(\Pi') \ge \adj^*_{\bar{\eps}} \ge 0$ by definition of $\alpha^*_{\bar{\eps}}$, and every cycle $\Gamma \in \mathcal{N}$ has $\adj_{\bar{\eps}}(\Gamma) \ge 0$ by the cycle invariant~\ref{I:cycle-invariant}.
Let $\Pi$ be an augmenting path with $\adj_{\ubar{\eps}}(\Pi) \le \adj^*_{\bar{\eps}}$.
Then
\[
\begin{aligned}
	\net\cost(\Pi) &\le \adj^*_{\bar{\eps}}
    \le \sum_{\Pi' \in \mathcal{P}} \adj_{\bar{\eps}}(\Pi') + \sum_{\Gamma \in \mathcal{N}} \adj_{\bar{\eps}}(\Gamma) &\\
    &= \sum_{\Pi' \in \mathcal{P}} \Paren{ \net\cost(\Pi') + c_0 \bar{\eps} \cdot \norm{\Pi'} }
        + \sum_{\Gamma \in \mathcal{N}} \Paren{ \net\cost(\Gamma) + c_0 \bar{\eps} \cdot \norm{\Gamma} } &\\
    &\le \cost(M_\opt) - \cost(M) + c_0 \bar{\eps} \cdot \Paren{ \cost(M_\opt) + \cost(M) } &\\
    &= (1 + c_0 \bar{\eps})\cdot \cost(M_\opt) + c_0 \bar{\eps} \cdot \cost(M) - \cost(M) &\\
    &\le \Paren{1 + c_0 \bar{\eps} \Paren{2 + \frac{\eps}{2}}}\cdot  \cost(M_\opt) - \cost(M)
        \; (\text{induction hypothesis}) \\
    &\le \Paren{1 + \frac{\eps}{8} \Paren{2 + \frac{\eps}{2}}}\cdot  \cost(M_\opt) - \cost(M)
        \; \Paren{\bar{\eps} = \frac{\eps}{c_1} \text{ and } c_1 \ge 8c_0} \\
    &\le \Paren{1 + \frac{\eps}{2}}\cdot  \cost(M_\opt) - \cost(M).
        \quad (0 \le \eps \le 1)
\end{aligned}
\]
Again, $\textsc{FindPath}$ returns an augmenting path $P$ with $\adj_{\ubar{\eps}}(P) \le \adj^*_{\bar{\eps}}$.
Therefore
\begin{align}
\cost(M \oplus P)
	&= \cost(M) + \net\cost(P)
    \le \cost(M) + \Paren{1+\frac{\eps}{2}} \cdot \cost(M_\opt) - \cost(M) \nonumber\\
	&= \Paren{1+\frac{\eps}{2}} \cdot \cost(M_\opt). \label{eq:path-cost2}
\end{align}
After augmenting $M$ with $P$, the algorithm may cancel a sequence of reducing cycles.
Since each of these cycles has negative net cost, the cycle cancellations only reduces the cost of the matching.
We conclude that the cost of an intermediate matching remains at most $(1+\frac{\eps}{2})\cdot \cost(M_\opt)$ during the $i$-th iteration.
\end{proof}

\begin{lemma}
\label{L:lengthsof}
The following two statements hold:
\begin{enumerate}[label=(\roman*)]\itemsep=0pt
    \item Let $M$ be an intermediate matching and $\Pi$ an augmenting path in $\vec{G}_M$ with $\adj_{\ubar{\eps}}(\Pi) \le \adj^*_{\bar{\eps}}$.
        Then, $\norm{\Pi} \le \frac{27\sqrt{d}n}{\eps}$.
    \item Let $M$ be an intermediate matching and $\Gamma$ a reducing cycle in $\vec{G}_M$.
        Then, $\norm{\Gamma} \le \frac{27\sqrt{d}n}{\eps}$.
\end{enumerate}
\end{lemma}

\FullVer{
\begin{proof}
Consider the matching $M$ at the beginning of an iteration, and an augmenting path $\Pi$ with $\adj_{\ubar{\eps}}(\Pi) \le \adj^*_{\bar{\eps}}$.
By applying Lemma~\ref{L:intermediate_augpath_full},
(\ref{P:opt_bounds}),
and $\eps < 1$,
\[
\begin{aligned}
\norm{\Pi} = \cost(\Pi)
    &\le \cost(M) + \cost(M \oplus \Pi)\\
	&\le \cost(M) + \Paren{1 + \frac{\eps}{2}} \cdot \cost(M_\opt) 
			\qquad \mbox{(By Eq.~(\ref{eq:path-cost1}) and (\ref{eq:path-cost2}))}\\
    & \le (2 + \eps) \cdot \cost(M_\opt)
    \le \frac{27\sqrt{d}n}{\eps}.
\end{aligned}
\]
If $M$ is an intermediate matching and $\Gamma$ is a reducing cycle, the same bound holds without applying Eq.~(\ref{eq:path-cost1}) and (\ref{eq:path-cost2}).
Instead, we can argue that $\cost(M \oplus \Pi) < \cost(M)$ since $\Pi$ is reducing:
\[ 
\norm{\Gamma} 
    = \cost(\Gamma) \le \cost(M) + \cost(M \oplus \Gamma)
	< 2 \cdot \cost(M) \le (2 + \eps) \cdot \cost(M_\opt) \\
    \le \frac{27\sqrt{d}n}{\eps}. 
\]
\end{proof}
}

We now analyze the running time of the algorithm.
Let $\Pi_1, \ldots, \Pi_n$ be the sequence of augmenting paths computed by the algorithm, let $\mathcal{N}_i$ be the set of alternating cycles that were canceled after augmenting by $\Pi_i$ during a call to the $\textsc{Augment}$ procedure, and let $M_i$ be the matching returned by the $\textsc{Augment}(\Pi_i, M_{i-1})$.
Then by Lemma~\ref{L:procedure_times}, the total time spent by the algorithm is $(\sum_i \abs{\Pi_i} + \sum_{\Gamma \in \mathcal{N}} \abs{\Gamma}) \cdot (\eps^{-1}\log n)^{O(d)}$, where $\mathcal{N} \coloneqq \bigcup_{i} \mathcal{N}_i$.
Since the cost of each edge in $G$ is at least 1 by (\ref{P:int_coords}), $\abs{\Pi_i} \le \norm{\Pi_i}$ and $\abs{\Gamma} \le \norm{\Gamma}$, so we will bound their Euclidean lengths instead.

We use the shorthand \EMPH{$\adj_{\theta, i}$} to denote $\adj_{\theta, M_i}$---the $\bar{\eps}$-adjusted cost after the $i$-th iteration where $M_i$ is a partial matching with $i$ edges---and set $\EMPH{$\adj^*_i$} \coloneqq \adj^*_{\theta, M_i}$.
We begin by bounding $\adj^*_i$.

\begin{lemma}
\label{L:adj-bound}
$\displaystyle \adj^*_i \le O\Paren{ \frac{n}{\eps(n-i)} }$.
\end{lemma}

\begin{proof}
$M_\opt \oplus M_i$ consists of a set of pairwise-disjoint alternating cycles and 
	$n-i$ augmenting paths $P_1, \dots, P_{n-i}$.
Since $\alpha^*_{\bar{\eps}, i}$ is the minimum $\bar{\eps}$-adjusted cost of an augmenting path in \smash{$\vec{G}_{M_i}$}, we have
\[
\begin{aligned}
(n-i) \cdot \adj^*_i
& \le \sum_{j=1}^{n-i} \adj_{\bar{\eps}, i}(P_j)
  \le \sum_{j=1}^{n-i} \Paren{ \net\cost_{M_i}(P_j)  + c_0\bar{\eps}\cdot \norm{P_j} } \\
& \le \sum_{j=1}^{n-i} \bigl( \cost(P_j \cap M_\opt) - \cost(P_j \cap M_i) \\
& + \frac{c_0 \eps}{c_1} \sum_{j=1}^{n-i} \Paren{ \cost(P_j \cap M_\opt) + \cost(P_j \cap M_i) } \bigr) \\
& \le \cost(M_\opt) + \frac{\eps}{8} \Paren{\cost(M_\opt) + \cost(M_i)}
	= O(n/\eps),
\end{aligned}
\]
where the last inequality follows from property (\ref{P:opt_bounds}) of the input and combining Lemma~\ref{L:intermediate_augpath_full}(i) with (\ref{P:opt_bounds}) and $\eps < 1$.
\end{proof}

 \begin{corollary}
 \label{C:adj_cost_bound}
 $\displaystyle \sum_{i=0}^{n-1} \adj^*_i = 
	 O(\eps^{-1} n \log n)$.
 \end{corollary}

\begin{lemma}
\label{L:sum_of_lengths_full}
$\displaystyle \sum_{i} \,\norm{\Pi_i} + \sum_{\Gamma \in \mathcal{N}} \,\norm{\Gamma} = 
	O(\eps^{-2}n\log^2 n)$.
\end{lemma}

\begin{proof}
Recall that $\textsc{Find-Path}$ guarantees
$\adj_{\ubar{\eps}, i}(\Pi_{i+1}) \le \adj^*_i$.
Therefore, by Lemma~\ref{L:adj-bound},
\[ 
	\sum_{i=1}^{n} \adj_{\ubar{\eps}, i-1}(\Pi_{i}) = O(\eps^{-1}n \log n).
\]
Since $\adj_{\ubar{\eps}}(\Gamma) < 0$ for all cycles $\Gamma \in \mathcal{N}$, we have
\[
\sum_{i=1}^{n} \adj_{\ubar{\eps}, i-1}(\Pi_{i}) + \sum_{\Gamma \in \mathcal{N}} \adj_{\ubar{\eps}}(\Gamma)
	= O(\eps^{-1}n \log n).
\]
On the other hand, by definition of the adjusted cost,
\begin{equation}
\label{E:sum_of_lengths1}
\begin{aligned}
\sum_{i=1}^{n} \adj_{\ubar{\eps}, i-1}(\Pi_{i}) + \sum_{\Gamma \in \mathcal{N}} \adj_{\ubar{\eps}}(\Gamma)
    &= \sum_{i=1}^{n} \net\cost(\Pi_i) + \sum_{\Gamma \in \mathcal{N}} \net\cost(\Gamma) \\
    &+ c_0 \ubar{\eps} \Paren{\sum_{i=1}^{n} \,\norm{\Pi_i} + \sum_{\Gamma \in \mathcal{N}} \,\norm{\Gamma}}.
\end{aligned}
\end{equation}
Observe that the augmentation of all paths and cycles results in the final matching $M_\alg$.
	Therefore, the net-cost terms in~(\ref{E:sum_of_lengths1}) sum to $\cost(M_\alg) \ge 0$, and we obtain
\[
\begin{aligned}
\sum_{i=1}^{n} \,\norm{\Pi_i} + \sum_{\Gamma \in \mathcal{N}} \,\norm{\Gamma}
    &= \frac{1}{c_0\ubar{\eps}} \Paren{ O\Paren{\frac{n\log n}{\eps}} - \cost(M_\alg)} \\
    &= \frac{c_1 c_2}{c_0} \cdot \frac{\log n}{\eps} \cdot O\Paren{\frac{n\log n}{\eps}}
    = O\Paren{\frac{n\log^2 n}{\eps^2}}.
\end{aligned}
\]
The second equality is obtained by substituting the value of $\ubar{\eps}$ and using $\cost(M_\alg) > 0$.
\end{proof}

\NotFullVer{Using (P1), (P2), and Lemmas~\ref{L:procedure_times} and~\ref{L:sum_of_lengths_full}, we obtain
}
\begin{lemma}
	\label{L:run-time}
	The overall algorithm runs in $n\cdot (\eps^{-1}\log n)^{O(d)}$ time.
\end{lemma}
\FullVer{
\begin{proof}
	By Lemma~\ref{L:procedure_times}, 
	$\textsc{Find-Path}$ takes $\abs{\Pi} \cdot (\eps^{-1}\log n)^{O(d)}$ time and 
	$\textsc{Augment}(\Pi, M)$ takes 
	$\bigl(\abs{\Pi} + \sum_{i} \abs{\Gamma_i}\bigr)\cdot (\eps^{-1}\log n)^{O(d)}$ time,  where $\Set{\Gamma_i}$ are the reducing cycles canceled by the procedure.
Let $\Pi_1, \ldots, \Pi_n$ be the sequence of augmenting paths computed by the algorithm,
	and let $\mathcal{N}$ be the set of alternating cycles that were canceled throughout the calls to the
	$\textsc{Augment}$ procedure.
The total time spent by the algorithm is $\bigl(\sum_i \abs{\Pi_i} + \sum_{\Gamma \in \mathcal{N}} \abs{\Gamma}\bigr) \cdot (\eps^{-1}\log n)^{O(d)}$.
	Since the cost of each edge in $G$ is at least 1 (by (P1) and (P2)),
	$\abs{\Pi_i} \le \norm{\Pi_i}$ and $\abs{\Gamma} \le \norm{\Gamma}$.
	The lemma now follows from Lemma~\ref{L:sum_of_lengths_full}.
\end{proof}
}

Lemmas~\ref{L:intermediate_augpath_full} and~\ref{L:run-time} together prove Theorem~\ref{Th:main_alg}.

\FullVer{
\paragraph{Remark.}
As mentioned in the introduction, the second term in the adjusted cost acts as a regularizer.
The Hungarian algorithm computes the minimum net-cost augmenting path at each step, which ensures that the residual graph has no negative cycles, but the total number of edges in augmenting paths can be as bad as $\Omega(n^2)$.
By adding the regularizer term, we ensure that the edge-length of the augmenting paths to be $O(\eps^{-2} n\log^2 n)$.
The residual graph may now contain negative cycles, but we ensure that the net-cost of these cycles is not too negative by enforcing that the $\bar{\eps}$-adjusted cost of a cycle to be non-negative 
through select reducing cycle cancellations.
}

\section{Proof of Expansion and Lifting Inequalities}
\label{S:ineq-proofs}

Finally, we prove the
expansion and lifting inequalities. 
We first present Lemma~\ref{L:expansion}, then two technical lemmas used for proving Lemma~\ref{L:lift}, and finally Lemma~\ref{L:lift} itself.

\subsection{Proof of expansion inequality}

\noindent\textbf{Lemma~\ref{L:expansion}.}
\textit{
    Let $\cell$ be a stable cell, and let $(X,Y) \in E_\cell$.
    Then
    \[
    \adj_{\ubar{\eps}}(\Phi_{\cell}(X, Y)) \le w_\cell(\pi_{\cell}(X,Y)).
    \]
    For $(X,Y) = (A_\subcell,B_{\subcell'})$, the inequality holds for all choices of its tips in $A_\subcell \times B_{\subcell'}$.
}

\begin{proof}
We write $\pi_\cell(X, Y)$ as its arc sequence $\Seq{\eta_1 \circ \eta_2 \circ \cdots \circ \eta_k}$.
	Let $\Pi \coloneqq \Phi(\eta_1) \circ \Phi(\eta_2) \circ \cdots \circ \Phi(\eta_k)$. Recall that if $\eta_i$ is an 
	internal arc of the form $(A_\subcell,B_{\subcell'})$ and $i>1$ (resp.\ $i<k$), then 
	the starting (resp.\ ending) tip of $\Phi(\eta_i)$ is changed to the ending (resp.\ starting) tip of 
	$\Phi(\eta_{i-1})$ (resp.\ $\Phi (\eta_{i+1}))$ during the concatenation. 

$\Pi$ is a possibly self-intersecting path in $\vec{G}_M$, and $\Phi_\cell(X,Y)$ is formed by simplifying $\Pi$ (cycle removal) as described in Section~\ref{S:expanding}.
Since $\cell$ is stable, none of these cycles were reducing.
Therefore $\adj_{\ubar{\eps}}(\Phi_\cell(X, Y)) \le \adj_{\ubar{\eps}}(\Pi)$.
We complete the proof by showing $\adj_{\ubar{\eps}}(\Pi) \le w_\cell(\pi_\cell(X,Y))$.

Recall that $\pnlty \coloneqq c_4 \ubar{\eps}$.
We choose $c_4 \ge 2 c_0 \sqrt{d}$.
For each arc $\eta_j \coloneqq (X_j, X_{j+1})$ in $\pi_\cell(X, Y)$, let $\subcell_j$ be the subcell containing $X_j$, 
	and $\cnter_j$ the center of $\subcell_j$.
	Recall that $\diam(\subcell_j) = \diam(\subcell_{j+1})\le \smash{\frac{\pnlty}{4}\ell_i}$.
We obtain a bound for each $\eta_j$:
\smallskip

	\noindent\textit{Case 1: $\eta_j$ is a {bridge arc}.}
	First suppose that $\eta_j$ is a non-matching arc, i.e., $X_j \subseteq A_{\subcell_j}, 
	X_{j+1} \subseteq B_{\subcell_{j+1}}$ and $\Phi(\eta_j)=(a_j,b_{j+1})$ for some pair in 
	$A_{\subcell_j}\times B_{\subcell_{j+1}}$.
    \[
    \begin{aligned}
        \adj_{\ubar{\eps}}(a_j, b_{j+1})
        &= \norm{a_j - b_{j+1}} + c_0 \ubar{\eps} \cdot \norm{a_j - b_{j+1}} \\
        &\le \norm{\cnter_j - \cnter_{j+1}} + 2 \cdot \smash{\textstyle\frac{\pnlty}{4}} \ell_i + c_0 \ubar{\eps} \smash{\sqrt{d}} \ell_i \\
        &\le \norm{\cnter_j - \cnter_{j+1}} + \pnlty\ell_i 
        = w_\cell(X_j, X_{j+1}).
    \end{aligned}
    \]
	Next, if $\eta_j$ is a matching  arc then
	$\Phi(\eta_j) = (b_j, a_{j+1}) \in M \cap B_{\subcell_j}\times A_{\subcell_{j+1}}$.
    We repeat a similar analysis on this single matching edge:
    \[
    \begin{aligned}
    \adj_{\ubar{\eps}}(\Phi(\eta_j))
	    &= -\norm{b_j - a_{j+1}} + c_0 \ubar{\eps} \cdot \norm{b_j - a_{j+1}}\\
	     &\le -\norm{\cnter_j - \cnter_{j+1}} + \pnlty\ell_i = w_\cell(X_j, X_{j+1}).
    \end{aligned}
    \]
    \smallskip

	\noindent\textit{Case 2:  $\eta_j=(X_j,X_{j+1})$ is an {internal arc}.}
	Let $(\cell',\Delta,\Delta')$ be the weight certificate of $\eta_j$, and let $X'_j \subset \Delta$ (resp.\ 
	$X'_{j+1} \subset\Delta')$ be the child cluster of $X_j$ (resp.\ $X_{j+1})$ in $\cell'$. Then 
	$\Phi(\eta_j)$ is the expansion $\Phi_{\cell'}(X'_j,X'_{j+1})$ with its starting and ending tips possibly 
	being modified to another point in $\subcell_j$ and $\subcell_{j+1}$, respectively.
%
    %
	Changing the tips increases the length of the first and last edges in $\Phi(\eta_j)$ by at most 
	$\diam(\subcell_j) = \diam(\subcell_{j+1})\le \smash{\frac{\pnlty}{4}\ell_i}$ each, thus
    \[
    \begin{aligned}
        \adj_{\ubar{\eps}}(\Phi(\eta_j))
        &\le \adj_{\ubar{\eps}}(\Phi_{\cell'}(X'_j, X'_{j+1})) + \frac{\pnlty}{2} \ell_i 
        \le w_\cell(X_j, X_{j+1}).
    \end{aligned}
    \]
The last inequality follows because $(\cell',\Delta,\Delta')$ is the arc-weight certificate of $\eta_j$.

Summing over the expansions of all arcs in $\Pi$, we obtain
$\adj_{\ubar{\eps}}(\Pi) \le w_\cell(\pi_\cell(B_{\subcell}, A_{\subcell'}))$.
\end{proof}

\FullVer{
\subsection{Intermediate lemmas for the lifting inequality}

\begin{lemma}
\label{L:cover-level}
Let $(p, q)$ be a pair of points in $A \cup B \subset \reals^d$.
Let $i$ be the smallest value for which a cell $\cell \in \Cover_i$ contains both $p$ and $q$.
Then
\begin{enumerate}[label=(\roman*)]
    \item $\ell_{i-2} \le \norm{p-q}_\infty \le \ell_i$ and
    \item $\ell_{i-2} \le \norm{p-q}_2 \le \sqrt{d} \ell_i$.
\end{enumerate}
\end{lemma}

\begin{proof}
Part~(ii) follows from (i). 
The upper bound in (i) follows because the side-length of $\cell$ is $\ell_i$.
Let $p = (p_1, \ldots, p_d)$ and $q = (q_1, \ldots, q_d)$.
Without loss of generality, assume that $q_j \ge p_j$ for all $j$.
Assume for contradiction that $\norm{p - q}_\infty < \ell_{i-2}$, then $q_j - p_j < \ell_{i-2}$ for all $1 \le j \le d$.
For $1 \le j \le d$, let $b_j = 1$ if $(p_j \mod \ell_{i-1}) \ge \ell_{i-2}$ and $0$ otherwise.
We claim that the (unique) cell $\cell'$ in the copy of $\Grid_{i-1}$ translated by $\ell_{i-2}(b_1, \ldots, b_d)$, i.e., in $\Grid_{i-1} + \ell_{i-2}(b_1, \ldots, b_d)$, that contains\footnotemark~$p$ also contains $q$.
Indeed, suppose $\cell' = [0, \ell_{i-1}]^d + (a_1 + b_1\ell_{i-2}, \ldots, a_d + b_d\ell_{i-2})$ for some $a_1, \ldots, a_d \in \ell_{i-1}\ints$.
By construction, $p_j - (a_j + b_j\ell_{i-2}) \le \ell_{i-2}$.
Therefore
\[
    q_j - (a_j + b_j\ell_{i-2}) = q_j - p_j + p_j - (a_j + b_j\ell_{i-2})
        < \ell_{i-2} + \ell_{i-2}
        < \ell_{i-1},
\]
or $q_j \le \ell_{i-1} + a_j + b_j\ell_{i-2}$, which implies $q \in [0, \ell_{i-1}]^d + (a_1 + b_1\ell_{i-2}, \ldots, a_d + b_d\ell_{i-2}) = \cell'$.
This contradicts the claim that $i$ is the lowest level for which a cell contains both $p$ and $q$.
Hence, $\norm{p - q}_\infty \ge \ell_{i-2}$.
\end{proof}

\footnotetext{
Here we assume that $p$ does not lie on the boundary of the grid.
If $p$ lies on the grid boundary then $\cell'$ is the cell that contains $p^+ = (p_1 + \eps_0, \ldots, p_d + \eps_0)$ where $\eps_0 > 0$ is some infinitesimally small value.
}

\begin{figure}
    \centering
    \includegraphics[width=0.23\textwidth, page=1]{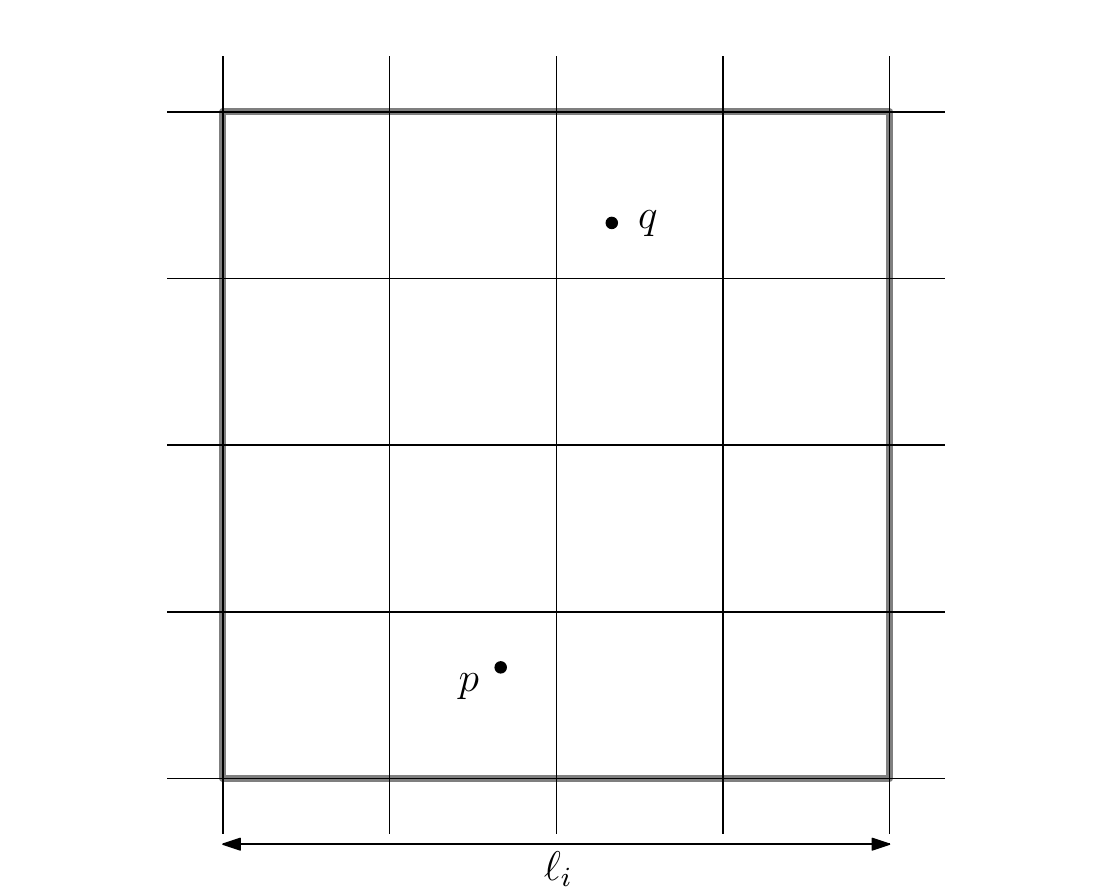}
    \includegraphics[width=0.23\textwidth, page=2]{lem_coverlevel}
    \caption{
        \textbf{Left}: Proof of Lemma~\ref{L:cover-level} upper bound.
        The level-$i$ cell in bold.
        Grid lines are those of $\Grid_{i-2}$.
        \textbf{Right}: Proof of Lemma~\ref{L:cover-level} lower bound.
        If $p$ and $q$ are too close in all dimensions, a lower-level cell (red) contains them both.
    }
    \label{}
\end{figure}

\begin{lemma}
\label{L:lift_single}
	Let $\cell$ be a level-$i$ cell and $(X,Y)\in E_\cell$ be a pair of clusters.  Let $\cell'$ be a 
	level-$i'$ ancestor cell of $\cell$ for $i' > i$, and $X'$ (resp.\ $Y'$) the ancestor cluster of $X$ (resp. $Y$) 
	in $\cell'$.  Then,
\[
    w_{\cell'}(\pi_{\cell'}(X', Y')) \le w_{\cell}(\pi_{\cell}(X, Y)) + 2\pnlty\ell_{i'}.
\]
\end{lemma}

\begin{proof}
Let \(
\cell = \cell_0, \cell_1, \ldots, \cell_t = \cell''
\)
be the sequence of parent-child cells starting from $\cell$ up to $\cell''$.
Define $X_j$ (resp.\ $Y_j$) to be the ancestor cluster of $X$ (resp.\ $Y$) in $\cell_j$.

For $j \ge 1$, the definition of the weight of internal edges~(\ref{E:weight_match_internal}) implies that
\begin{equation}
\label{E:singlelift1}
\begin{aligned}
    w_{\cell_j}(\pi_{\cell_j}(X_j, Y_j))
        &\le w_{\cell_j}(X_j, Y_j) \\
	&\le \adj_{\ubar{\eps}}(\Phi_{\cell_{j-1}}(X_{j-1},Y_{j-1})) + \pnlty\ell_j \\
        &\le w_{\cell_{j-1}}(X_{j-1},Y_{j-1}) + \pnlty\ell_j,
\end{aligned}
\end{equation}
where the last inequality follows from Lemma~\ref{L:expansion}.
Using (\ref{E:singlelift1}) recursively, we obtain
\[
\begin{aligned}
    w_{\cell'}(\pi_{\cell'}(X', Y'))
        &= w_{\cell_t}(\pi_{\cell_t}(X_t,Y_t)) \\
	&\le w_{\cell_{t-1}}(\pi_{\cell_{t-1}}(X_{t-1}, Y_{t-1})) + \pnlty\ell_{i'} \\
        &\le w_{\cell}(\pi_{\cell}(X,Y)) + \pnlty \sum_{j=1}^{t} \frac{\ell_{i'}}{2^{t-j}} \\
        &\le w_{\cell}(\pi_{\cell}(X,Y)) + 2\pnlty\ell_{i'}.  
\end{aligned}
\vspace{-12pt}
\]
\end{proof}

\noindent\textbf{Lemma~\ref{L:container_exists}.}
\textit{Let $\Pi$ be either a reducing cycle, or an  augmenting path in $\vec{G}_M$ with the minimum $\bar{\eps}$-adjusted cost.
Then there exists a level $i \in \set{0,\ldots,h}$ and $\cell \in \Cover_i$ such that $\Pi$ lies completely in $\cell$.
}
%
\begin{proof}
For either option of $\Pi$ from the lemma statement (reducing cycle, low-adjusted-cost augmenting path), Lemma~\ref{L:lengthsof} implies that
\(
    \norm{\Pi} \le \frac{27\sqrt{d}n}{\eps}.
\)
For $1 \le j \le d$, let 
\[
    p^-_j \coloneqq \min_{p \in \Pi} x_j(p) 
	\quad \mbox{and} \quad 
	p^+_j \coloneqq \max_{p \in \Pi} x_j(p).
\]
Then, $p^+_j - p^-_j \le \norm{\Pi}$.
Let $p^- = (p^-_1, \ldots, p^-_d)$ and $p^+ = (p^+_1, \ldots, p^+_d)$; $\norm{p^+ - p^-}_\infty \le \norm{\Pi}$.
	If $i$ is the value of the lowest level for which a cell $\cell \in \Cover_i$ contains both $p^-$ and $p^+$, then
	by Lemma~\ref{L:cover-level}, $\ell_{i-2} \le \norm{p^+-p^-}\le \norm{\Pi}$, or  
	$i \le 2 + {\Ceil{\Big.\log \Paren{ {27\sqrt{d}n}/{\eps} }}}$.

Let $\mathsf{B} \coloneqq \smash{\prod_{j=1}^d [p^-_j, p^+_j]}$ be the axis-aligned hypercube determined by $p^-$ and $p^+$.
Then, $\Pi \subseteq \mathsf{B} \subseteq \cell$.
Assuming $c_3$ is chosen sufficiently large and $\e > 1/n$ so that $h = c_3 \Ceil{\log n} \ge 2 + 
		\Ceil{\Big.\log \Paren{ {27\sqrt{d}n}/{\eps} }}$, the lemma holds.
\end{proof}
}

\subsection{Proof of lifting inequality}

\NotFullVer{
	We first state the two lemmas needed to prove the lifting inequality.
\begin{lemma}
\label{L:cover-level}
Let $(p, q)$ be a pair of points in $A \cup B \subset \reals^d$.
Let $i$ be the smallest value for which a cell $\cell \in \Cover_i$ contains both $p$ and $q$.
Then
\begin{enumerate}[label=(\roman*)]
    \item $\ell_{i-2} \le \norm{p-q}_\infty \le \ell_i$ and
    \item $\ell_{i-2} \le \norm{p-q}_2 \le \sqrt{d} \ell_i$.
\end{enumerate}
\end{lemma}

\begin{lemma}
\label{L:lift_single}
	Let $\cell$ be a level-$i$ cell and $(X,Y)\in E_\cell$ be a pair of clusters.  Let $\cell'$ be a 
	level-$i'$ ancestor cell of $\cell$ for $i' > i$, and $X'$ (resp.\ $Y'$) the ancestor cluster of $X$ (resp. $Y$) 
	in $\cell'$.  Then,
\[
    w_{\cell'}(\pi_{\cell'}(X', Y')) \le w_{\cell}(\pi_{\cell}(X, Y)) + 2\pnlty\ell_{i'}.
\]
\end{lemma}
We are now ready to prove the lifting inequality.
}

\noindent\textbf{Lemma~\ref{L:lift}.}
\textit{Let $\Pi$ be an alternating path in $\vec{G}_M$ from a point $p$ to a point $q$ (possibly $p=q$ in the case $\Pi$ is a cycle) that is completely contained in a cell of level $\le h$.
Let $\cell$ be a cell at the smallest level that contains $\Pi$, and
let $X$ (resp.\ $Y$) be the cluster in $V_\cell$ containing $p$ (resp.\ $q$).
Then either $\cell$ is not stable or
\(
	w_\cell(\pi_\cell(X, Y)) \le \adj_{\bar{\e}}(\Pi).
\)
}

\begin{proof}
If $\cell$ is not stable then the lemma follows trivially, so assume that $\cell$ is stable, which implies that all descendants of $\cell$ are also stable.
Namely, arc weights and expansions are well-defined for every descendant of $\cell$.
We construct a compressed path $P$ in $G_\cell$ where
	\begin{equation}
		\label{eq:c-path-length}
    w_\cell({P}) \le \net\cost(\Pi) + c_5 i \pnlty\norm{\Pi}
	\end{equation}
for some constant $c_5 \ge 64$.
If $\cell$ is at level $i$, then by assumption $i \le h \le c_3\log n$, and since $\pnlty = c_4 \ubar{\eps}$, $\ubar{\eps} = \frac{\bar{\eps}}{c_2\log n}$, we obtain
\begin{equation}
\begin{aligned}
    w_\cell(P)
        &\le \net\cost(\Pi) + c_5 (c_3 \log n) c_4 \Paren{\frac{\bar{\eps}}{c_2\log n}} \cdot \norm{\Pi} \\
        &\le \net\cost(\Pi) + c_0 \bar{\eps} \cdot \norm{\Pi}
        = \adj_{\bar{\eps}}(\Pi)
\end{aligned}
\end{equation}
provided that $c_0 \ge c_3 c_4 c_5 / c_2$.

	We prove (\ref{eq:c-path-length}) by induction over hierarchy levels.
Let $\cell$ be the smallest cell containing $\Pi$ and let $i$ be the level of $\cell$.
If $i=0$ then $\Pi$ is an empty path, as each cell at level~0 contains no edge of $\vec{G}_M$, so the lemma holds trivially.
Assume that the lemma holds for all levels less than $i$.

Let $\Pi = \langle p_1 = p, p_2, \ldots, p_t = q \rangle$.
We first construct a sequence $P$ of nodes in $V_\cell$ by traversing $\Pi$ and using a greedy approach, and then refine it to ensure that the resulting sequence is a path in $G_\cell$.
For a point $p_k$, let $X(p_k)$ be the cluster in $V_\cell$ containing $p_k$.
To start, set ${P} \coloneqq \langle X_1\rangle = \langle X(p_1)\rangle$.
Suppose we have traversed a prefix of $\Pi$, constructed a prefix of ${P} = \langle X_1, X_2, \ldots, X_j \rangle$, and we are currently at point $p_k$ where $X(p_k) = X_j$.
Let $p^*$ be the furthest point of $\Seq{p_k, p_{k+1}, \ldots, p_t}$ such that the subpath $\Seq{p_k, \ldots, p^*} \subseteq \Pi$ lies in a single child cell of $\cell$.
This can be done by monitoring the set of children cells that cover the current prefix (initially the $\le \smash{2^d}$ children that cover $p_k$) and shrinking as we go.
There are a few special cases where $p^*$ may not be well-defined above.
\begin{itemize}
\item
If $p^* = p_k$ (the longest single-child prefix is one point), then $(p_k, p_{k + 1})$ corresponds to a bridge arc in $G_\cell$, and we set $p^* \from p_{k + 1}$.
\item
If the longest single-child prefix starting from $p_k$ is at least two points, but one child cell contains the entire suffix, we set $p^* \from q$ instead.
\end{itemize}
After setting $p^*$, we choose $X_{k+1} \coloneqq X(p^*)$.
If $p^*$ is the last point of $\Pi$, we stop.
Otherwise, we continue.

We note that $P$ is not necessarily a path in $G_\cell$ because it might contain a consecutive pair $X_j, X_{j+1}$ where $X_j$ and $X_{j+1}$ are both $A$-clusters or both $B$-clusters (hence, $(X_j,X_{j+1})$ is not an arc in the bipartite graph $G_\cell$).
We fix these pairs by inserting an extra cluster in between each such occurrence.
For each $X_j$, let $u_j = k$ be the index for which $X(p_k) = j$.
Let $\Pi_j = \langle p_{u_j}, \ldots, p_{u_{j+1}} \rangle \subseteq \Pi$.
	Assume both $X_j$ and $X_{j+1}$ are $A$-clusters (the other case can be handled in an analogous way),
then $p_{u_j}, p_{u_{j+1}} \in A$ and $\Pi_j \cap B$ is nonempty.
Choose any point $y \in \Pi_j \cap B$, set $Y_j = X(y)$, and replace the substring $X_j, X_{j+1}$ in $P$ with 
	$X_j, Y_j, X_{j+1}$.
Both $(X_j, Y_j)$ and $(Y_j, X_{j+1})$ are internal arcs in $G_\cell$, as a single child of $\cell$ contains all of $\Pi_j$, 
	and the sequence alternates between $A$- and $B$-clusters.

Let $P \coloneqq \Seq{X_1=X, X_2, \ldots, X_{s-1}, X_s =Y}$ be the resulting compressed path from $X=X(p)$ to $Y=X(q)$ 
in $G_\cell$.  The above construction guarantees that for any $j \le s-3$, no contiguous quadruple 
$X_j, X_{j+1}, X_{j+2}, X_{j+3}$ lie in a single child cell of $\cell$.
For $1 \le j \le s$, let $\subcell_j$ be the subcell of $\cell$ containing $X_j$, and let $\cnter_j$ be the center of $\subcell_j$.
We now bound $w_\cell({P})$ by relating $w_\cell(X_j,X_{j+1})$ to the cost of $\Pi_j$.
There are two main cases:
\smallskip

\noindent\textit{Case 1: $(X_j, X_{j+1})$ is a \textbf{bridge arc}.}
    Since $\diam(\subcell_j) = \diam(\subcell_{j+1}) \le \frac{\pnlty\ell_i}{4}$, we have
	\[
	\norm{p_{u_j} - p_{u_{j+1}}} - \frac{\pnlty\ell_i}{2} \le
	\norm{\cnter_j - \cnter_{j+1}} \le \norm{p_{u_j} - p_{u_{j+1}}} + \frac{\pnlty\ell_i}{2}.
	\]
	If $(X_j,X_{j+1})$ is a non-matching bridge arc, then
    \begin{equation}
    \label{E:lifting_2}
    \begin{aligned}
        w_\cell(X_j, X_{j+1})
        &= \norm{\cnter_j - \cnter_{j+1}} + \pnlty\ell_i \\
        &\le \norm{p_{u_j} - p_{u_{j+1}}} + \frac{\pnlty\ell_i}{2} + \pnlty\ell_i
        = \net\cost(\Pi_j) + \frac{3}{2} \pnlty\ell_i.
    \end{aligned}
    \end{equation}
    If $(X_j, X_{j+1})$ is a matching bridge arc, following a similar argument, we obtain
    \begin{equation}
    \label{E:lifting_2a}
    \begin{aligned}
        w_\cell(X_j, X_{j+1})
        &= -\norm{\cnter_j - \cnter_{j+1}} + \pnlty\ell_i \\
        &\le -\norm{p_{u_j} - p_{u_{j+1}}} + \frac{3}{2} \pnlty\ell_i
        = \net\cost(\Pi_j) + \frac{3}{2} \pnlty\ell_i.
    \end{aligned}
    \end{equation}
    \smallskip

\noindent\textit{Case 2: $(X_j, X_{j+1})$ is an \textbf{internal arc}.}
$(X_j, X_{j+1})$ is a pair of clusters contained in a single child cell $\cell'$ of $\cell$.
    Let $X'_j$ (resp.\ $X'_{j+1}$) be the child cluster of $X_j$ (resp.\ $X_{j+1}$) in $\cell'$ 
    that contains $p_{u_j}$ (resp.\ $p_{u_{j+1}}$). By the definition of internal arc weights, we have
    \begin{equation}
    \label{E:lifting_3a}
    \begin{aligned}
        w_\cell(X_j, X_{j+1})
	    &\le \adj_{\ubar{\eps}}(\Phi_{\cell'}(X'_j,X'_{j+1})) + \pnlty\ell_i \\
	    & \le w_{\cell'}(\pi_{\cell'}(X'_j, X'_{j+1})) + \pnlty\ell_i,
    \end{aligned}
    \end{equation}
    where the last inequality follows from Lemma~\ref{L:expansion} and the fact that $\cell'$ is stable.

    Let $\hat{\cell}$ be the smallest descendant of $\cell'$ that contains $\Pi_j$, and let $\hat{i} \le i-1$ be the level of $\hat{\cell}$ (note that $\hat{\cell}$ may be $\cell'$ itself).
    Let $\hat X_j$ (resp.\ $\hat{X}_{j+1}$) be the descendant cluster of $X_j$ (resp.\ $X_{j+1}$) in 
    $\hat{\cell}$ containing $p_{u_j}$ (resp.\ $p_{u_{j+1}}$).
    Since $\pi_{\hat{\cell}}(\hat{X}_j, \hat{X}_{j+1})$ is the minimum-weight path from 
    $\hat{X}_j$ to $\hat{X}_{j+1}$ in $G_{\hat{\cell}}$, by the induction hypothesis,
    \[
	    w_{\hat{\cell}}(\pi_{\hat{\cell}}(\hat{X}_j, \hat{X}_{j+1})) \le \net\cost(\Pi_j) + c_5 \hat{i} \pnlty \norm{\Pi_j}.
    \]
    Using Lemma~\ref{L:lift_single}, we obtain
    \begin{equation}
    \label{E:lifting_3b}
    \begin{aligned}
        w_{\cell'}(\pi_{\cell'}(X'_j, X'_{j+1}))
        &\le w_{\hat{\cell}}(\pi_{\hat{\cell}}(\hat{X}_j, \hat{X}_{j+1})) + 2\pnlty\ell_{i-1} \\
        &\le \net\cost(\Pi_j) + c_5 \hat{i} \pnlty \norm{\Pi_j} + \pnlty\ell_{i} \\
        &\le \net\cost(\Pi_j) + c_5 (i-1) \pnlty \norm{\Pi_j} + \pnlty\ell_i.
    \end{aligned}
    \end{equation}
    Substituting~(\ref{E:lifting_3b}) into~(\ref{E:lifting_3a}) we obtain
    \begin{equation}
    \label{E:lifting_3}
        w_\cell(X_j, X_{j+1})
        \le \net\cost(\Pi_j) + c_5 (i-1) \pnlty \norm{\Pi_j} + 2\pnlty\ell_i.
    \end{equation}

Combining (\ref{E:lifting_2}) and~(\ref{E:lifting_3}), the terms from~(\ref{E:lifting_3}) dominate, so
\begin{equation}
\label{E:lifting_4}
\begin{aligned}
    w_\cell({P}) &= \sum_{j = 1}^{s-1} w_\cell(X_j, X_{j+1}) \\
    &\le \sum_{j = 1}^{s-1} \Paren{ \net\cost(\Pi_j) + c_5 (i-1) \pnlty \norm{\Pi_j} + 2\pnlty\ell_i } \\
    &= \net\cost(\Pi) + c_5 (i-1) \pnlty \norm{\Pi} + 2(s-1)\pnlty\ell_i
\end{aligned}
\end{equation}

Recall that no contiguous quadruple $X_j, X_{j+1}, X_{j+2}, X_{j+3}$ (alternatively, three consecutive arcs of $P$) 
lie in a single child cell, for all $j \le s-3$.
Then, by Lemma~\ref{L:cover-level}, $\norm{\Pi_j \circ \Pi_{j+1} \circ \Pi_{j+2}} \ge \ell_{i-1}/4 = \ell_i/8$.
Summing,
\[
	\sum_{j=1}^{s-3} \,\norm{\Pi_j \circ \Pi_{j+1}\circ \Pi_{j+2}}
    \ge \Paren{s-3} \frac{\ell_i}{8}.
\]
In this sum, each subpath $\Pi_j$ is being counted at most three times.
Hence, the left-hand quantity is at most $3\norm{\Pi}$, therefore
\begin{equation}
\label{E:lifting_5}
    (s-3) \ell_i \le 24 \norm{\Pi}.
\end{equation}
Since $\cell$ is the smallest cell to contain $\Pi$, by Lemma~\ref{L:cover-level}, $\norm{\Pi} \ge \ell_i/4$.
Plugging (\ref{E:lifting_5})~into~(\ref{E:lifting_4}),
\[
\begin{aligned}
    w_\cell({P}) &\le \net\cost(\Pi) + c_5(i-1)\pnlty\norm{\Pi} + 2(s-1)\pnlty\ell_i \\
    &\le \net\cost(\Pi) + c_5(i-1)\pnlty\norm{\Pi} + 48\pnlty\norm{\Pi} + 4\pnlty\ell_i \\
    &\le \net\cost(\Pi) + c_5(i-1)\pnlty\norm{\Pi} + 64\pnlty\norm{\Pi} \\
    &\le \net\cost(\Pi) + c_5 i \cdot \pnlty\norm{\Pi},
\end{aligned}
\]
provided that $c_5 \ge 64$.
This completes the proof of the lemma.
\end{proof}




\FullVer{
\section{Preprocessing Step}
\label{S:preprocess}

We now describe the preprocessing step that makes the input ``well-conditioned'' at a slight increase in the cost of optimal matching.  It consists of two stages.

\mypara{A coarse approximation.}
The first stage computes an $O(n^2)$ approximation of $\cost(M_\opt)$, as follows:
Following the algorithm of Callahan-Kosaraju~\cite{ck-faggp-1993} (see also Har-Peled~{\cite[Chapter~4]{har-gaa-2011}})
we can construct in $O(n\log n)$ time a weighted graph $H$ on $A \cup B$ with edge weights $w_H(p, q) = \norm{p-q}$, such that for any pair of points $(p, q)$ in $A \cup B$, there is a path $P$ in $H$ from $p$ to $q$ such that $w_H(P) \le 2 \norm{p-q}$.
(In other words, graph $H$ is a \emph{Euclidean 2-spanner} of $A \cup B$.)
Next, we execute Kruskal's minimum-spanning tree algorithm on $H$ until each connected component $X$ of the forest has equal number of points of $A$ and $B$. 
Let $\GLOS[coarse-approx-opt]{Coarse approximation to cost of optimal matching}{w_0}$ be the weight of the last edge added by the algorithm.

\begin{lemma}
\label{L:coarse_approx}
$\frac{w_0}{2} \le \cost(M_\opt) < n^2 \cdot w_0$.
\end{lemma}

\begin{proof}
Let $e_0$ be the last edge added by the algorithm that connected two clusters $X_1$ and $X_2$ and merged them into a single cluster $X$.
By assumption, $\abs{X \cap A} = \abs{X \cap B}$ but $\abs{X_i \cap A} \neq \abs{X_i \cap B}$ for $i \in \set{1,2}$ (because otherwise the algorithm would have stopped sooner).
Suppose $\abs{X_1 \cap A} > \abs{X_1 \cap B}$.
Then there is an edge $(p,q) \in M_\opt$ whose one endpoint is in $A \cap X_1$ and another endpoint in $B \setminus X_1$.
By construction of~$H$, there is a path $\pi$ in $H$ from $p$ to $q$ such that $w_H(\pi) \le 2\norm{p-q}$.
The path must contain an edge $e' = (p',q')$ such that $p' \in X_1$ to $q' \not\in X_1$.
Since the algorithm chose to add $e_0$ instead of $(p', q')$, we have
\[
    w_0 = \norm{e_0} \le \norm{p'-q'} \le w_H(\pi) \le 2\norm{p-q} \le 2\cost(M_\opt).
\]
Hence, $\cost(M_\opt) \ge w_0/2$.

Next, let $X_1, \ldots, X_k$ be the connected components of the forest maintained by Kruskal's algorithm after $e_0$ was added.
For each $X_i$, let $M_i$ be an arbitrary perfect matching between $A \cap X_i$ and $B \cap X_i$.
By triangle inequality, the distance between any two points of $X_i$ (and in particular, any matching edge) is at most $(\abs{X_i}-1) w_0 < n w_0$.
Hence, $\cost(M_\opt) \le \cost(\bigcup_{i} M_i) \le n^2 \cdot w_0$, as claimed.
\end{proof}

\begin{figure}
    \centering
    \includegraphics[width=0.23\textwidth, page=1]{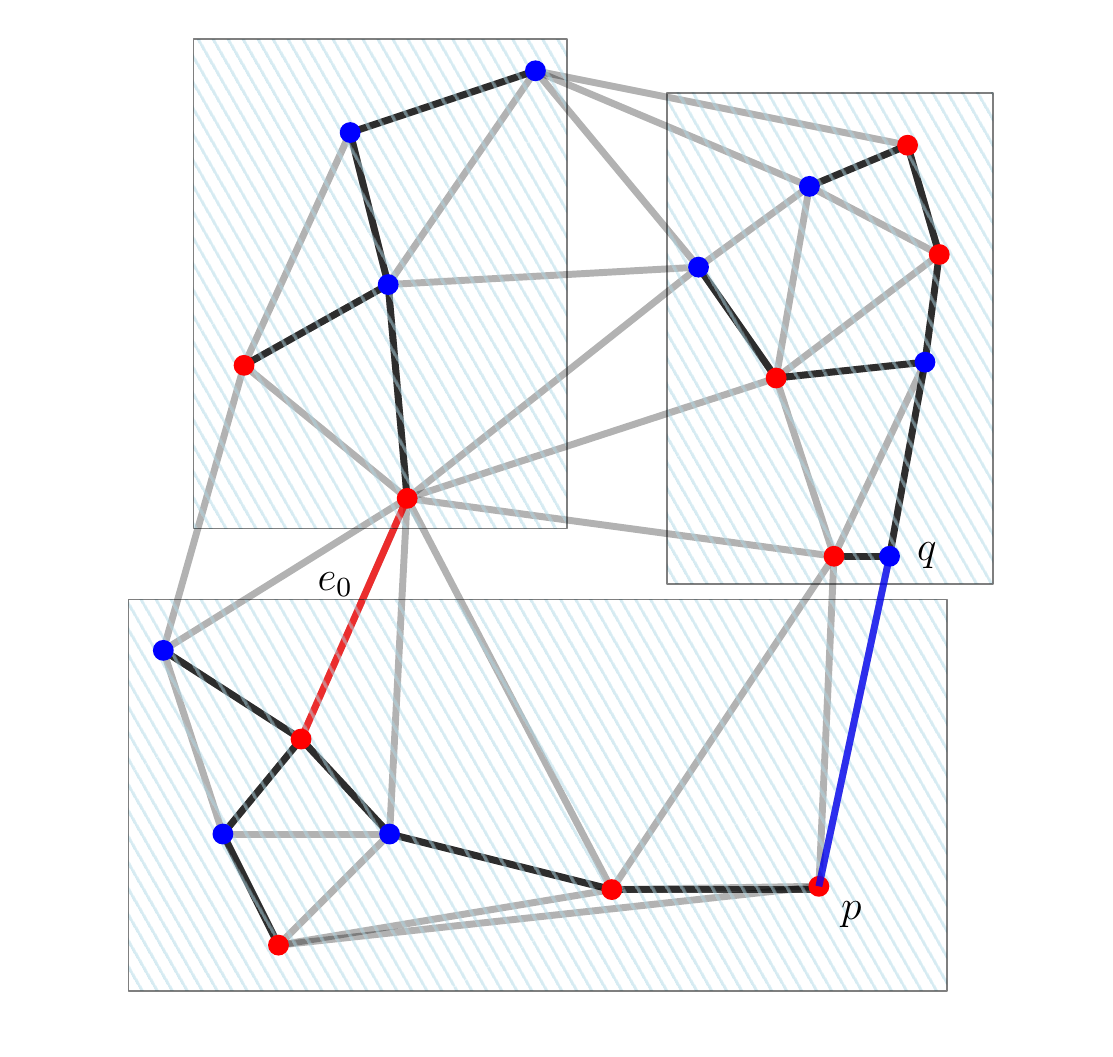}
    \includegraphics[width=0.23\textwidth, page=2]{lem2-1}
    \caption{
        Proof of Lemma~\ref{L:coarse_approx}.
        Edges of the 2-spanner ($H$) are in gray, the included MST edges are in black, shaded regions denote connected components.
        The last edge added, $e_0$, is in red.
        \textbf{Left}: Partitions reflect the state right before $e_0$ is added.
        $(p, q) \in M_\opt$ (blue) crosses out of the bottom partition.
        \textbf{Right}: $e_0$ is the longest MST edge out of all included edges, and component diameter is bounded by applying triangle inequality.
    }
    \label{fig:lem_coarse_approx}
\end{figure}

For any integer $i \in [-1, 2\log n]$\footnotemark, define $\beta_i \coloneqq 2^i \cdot w_0$.
By Lemma~\ref{L:coarse_approx}, there is an $i$ such that $\beta_{i-1} \le \cost(M_\opt) \le \beta_i$.
We run our algorithm for at most $c_0 \cdot n \polylog n$ steps, where $c_0$ is a sufficiently large constant specified below, on each choice of $\beta_i$.
In the $i$-th iteration, either the algorithm returns a perfect matching of $A$ and $B$ or terminates without computing a perfect matching.
Among the perfect matchings computed by the algorithm, we return the one with the smallest cost.
Theorem~\ref{Th:main_alg} ensures that if $\beta_{i-1} \le \cost(M_\opt) \le \beta_i$ then the algorithm returns a perfect matching of cost at most $(1+\eps) \cdot \cost(M_\opt)$.
Now forward, we assume that we have computed a value $\beta > 0$ such that
\begin{equation}
\label{E:costest}
    \cost(M_\opt) \le \GLOS[fine-approx-opt]{Constant approximation to cost of optimal matching}{\beta} \le 2 \cdot \cost(M_\opt).
\end{equation}

\footnotetext{
unless specified otherwise, all the logarithms are of base 2 throughout this paper
}

\mypara{Conditioning the input.}
We multiply the coordinates of each point by a factor of $\smash{\frac{8\sqrt{d}n}{\eps \beta}}$.
By (\ref{E:costest}), the cost of the optimal matching now lies in the range $\Brack{ \frac{4\sqrt{d}n}{\eps}, \frac{8\sqrt{d}n}{\eps} }$.
Next, we move each input point to its nearest integer grid point.
Since each point is moved a distance of at most $\smash{\frac{\sqrt{d}}{2}}$, the cost of optimal matching of perturbed points increases by $\frac{\sqrt{d}}{2} n \le \frac{\eps}{8} \cost(M_\opt)$.
If a grid point contains both red and blue points, we match them and delete them --- this is optimal by triangle inequality.
Hence, we assume each grid point contains points of at most one color.
Note that multiple points of one color may map to the same grid point, so we have a multiset of points.
Abusing the notation a little, we treat each copy as a separate point and let $A$ and $B$ denote the resulting (multi-)sets of perturbed points.
By construction,
the cost of optimal matching on the perturbed points $\cost(M_\opt)$ lies in
\[
\Brack{ \frac{4\sqrt{d}n}{\eps} - \frac{\sqrt{d}n}{2}, \frac{8\sqrt{d}n}{\eps} + \frac{\sqrt{d}n}{2} }
\subseteq \Brack{ \frac{3\sqrt{d}n}{\eps}, \frac{9\sqrt{d}n}{\eps} }
\]
since we assume $\eps \le 1$.

After this preprocessing step, which takes $O(n \log^2 n)$ time in total for any fixed $d$, we have point sets 
$A$ and $B$ that satisfy the three properties (\ref{P:int_coords})--(\ref{P:opt_bounds}) stated in Section~\ref{S:overview}.
}

\section{Concluding Remarks}
\label{S:conclude}

We conclude the paper with the following two important open questions:
(i) find a deterministic algorithm for $\eps$-optimal transport, and 
(ii) find an $O(n \poly\log n)$-time algorithm for RMS matching.  
While we circumvent the use of probabilistic embeddings, our new ideas still strongly rely on the use of triangle inequalities (in particular, in the proof of Lemmas~\ref{L:expansion} and~\ref{L:lift}).




\bibliographystyle{abbrv}
\bibliography{det-matching,extra}

\end{document}